\documentclass[journal]{IEEEtran}

\usepackage{mathtools}
\usepackage{amsmath} 
\usepackage{amssymb}
\usepackage{amsfonts}
\usepackage{verbatim} 
\usepackage{bm,color,soul}
\usepackage{algorithm,algorithmic}
\usepackage{cite}
\usepackage{caption}
\usepackage{subfigure}
\usepackage{stfloats} 
\usepackage[colorlinks, linkcolor=red, anchorcolor=blue, citecolor=green]{hyperref} 
\usepackage{graphicx} 
\usepackage{subeqnarray}
\usepackage{cases}
\usepackage{makecell} 
\usepackage{enumerate}
\usepackage{array}
\usepackage{url}
\usepackage{balance}

\newtheorem{rem}{Remark}
\newtheorem{prop}{Proposition}
\newtheorem{lem}{Lemma}

\newcommand{\rnum}[1]{\uppercase\expandafter{\romannumeral #1\relax}}
\ifCLASSINFOpdf
\else
\fi
\makeatletter
\renewcommand{\maketag@@@}[1]{\hbox{\m@th\normalsize\normalfont#1}}
\makeatother
\allowdisplaybreaks 

\IEEEoverridecommandlockouts

\begin{document}
%
\title{Multi-Active-IRS-Assisted Cooperative Sensing: Cram\'{e}r-Rao Bound and Joint Beamforming Design\vspace{-0.0em} }

\author{\IEEEauthorblockN{Yuan Fang,~\IEEEmembership{Member,~IEEE},  Xianghao Yu,~\IEEEmembership{Senior Member,~IEEE},\\ Jie Xu,~\IEEEmembership{Senior Member,~IEEE}, and Ying-Jun Angela Zhang,~\IEEEmembership{Fellow, IEEE}
	}\vspace{-0.8cm}
	\thanks{Part of this paper has been submitted to the 2024 IEEE Global Communications Conference \cite{fang2024active}. }
	\thanks{Y. Fang and X. Yu are with the Department of Electrical Engineering, City University of Hong Kong, Hong Kong, China (e-mail: yuanfang@cityu.edu.hk, alex.yu@cityu.edu.hk).}
	\thanks{J. Xu is with the School of Science and Engineering (SSE) and the Future Network of Intelligence Institute (FNii), The Chinese University of Hong Kong (Shenzhen), Guangdong, 518172, China (e-mail: xujie@cuhk.edu.cn).}
	\thanks{Y.-J. Zhang is with the Department of Information Engineering, The Chinese University of Hong Kong, Hong Kong, China (e-mail: yjzhang@ie.cuhk.edu.hk).}
	\thanks{X. Yu and J. Xu are the corresponding authors.}
}

\maketitle
\begin{abstract}
	This paper studies the multi-intelligent reflecting surface (IRS)-assisted cooperative sensing, in which multiple active IRSs are deployed in a distributed manner to facilitate multi-view target sensing at the non-line-of-sight (NLoS) area of the base station (BS). Different from prior works employing passive IRSs, we leverage active IRSs with the capability of amplifying the reflected signals  to overcome the severe multi-hop-reflection path loss in NLoS sensing. In particular, we consider two sensing setups without and with dedicated sensors equipped at active IRSs. In the first case without dedicated sensors at IRSs, we investigate the cooperative sensing at the BS, where the target's direction-of-arrival (DoA) with respect to each IRS is estimated based on the echo signals received at the BS. In the other case with dedicated sensors at IRSs, we consider that each IRS is able to receive echo signals and estimate the target's DoA with respect to itself. For both sensing setups, we first derive the closed-form Cram\'{e}r-Rao bound (CRB) for estimating target DoA. Then, the (maximum) CRB is minimized by jointly optimizing the transmit beamforming at the BS and the reflective beamforming at the multiple IRSs, subject to the constraints on the maximum transmit power at the BS, as well as the maximum amplification power and the maximum power amplification gain constraints at individual active IRSs. To tackle the resulting highly non-convex (max-)CRB minimization problems, we propose two efficient algorithms to obtain high-quality solutions for the two cases with sensing at the BS and at the IRSs, respectively, based on alternating optimization, successive convex approximation, and semi-definite relaxation. Finally, numerical results are provided to verify the effectiveness of our proposed design and the benefits of multi-active-IRS-assisted sensing compared to its counterpart with passive IRSs.
\end{abstract}


%
\IEEEpeerreviewmaketitle

\vspace{-3mm}
\section{Introduction}
\vspace{-1mm}
Integrating sensing into wireless communication systems has emerged as a prominent application scenario for future sixth-generation (6G) wireless networks \cite{cui2021integrating,zhang2021overview}. This catalyzes extensive emerging technologies, including autonomous driving, virtual reality, and airspace supervision. With the wireless sensing capability, cellular base stations (BSs) can extract useful environmental and object information from echo signals \cite{liu2022integrated}. Furthermore, BSs with massive antennas in communication networks are able to provide ultra-resolution and high-accuracy sensing. However, wireless sensing generally depends on the line-of-sight (LoS) channel between the BS and target, which is highly likely to be blocked by various infrastructures, vehicles, or vegetation in complex wireless environments, thus seriously limiting the sensing performance \cite{huang2019reconfigurable,hua2022joint,hua2023intelligent}.

The development of intelligent reflecting surfaces (IRSs) provides a viable solution to bypassing the blockage. This is achieved by constructing virtual LoS paths via reflecting the incident signals with properly controlled phases \cite{huang2019reconfigurable,di2020smart,wu2021intelligent}. In the literature, there have been various existing works aiming to enhance communication or sensing  capabilities by deploying {\it passive} IRSs in wireless systems \cite{song2024cramer}. Specifically, the authors in \cite{aubry2021reconfigurable} developed a novel received signal power model for IRS-assisted radar to detect the target located in the non-line-of-sight (NLoS) area. Receiving sensors were proposed to be installed on passive IRSs for target sensing and parameter estimation at the IRS \cite{shao2022target}, whereas IRS beamforming was optimized to maximize the sensing signal-to-noise ratio (SNR). The authors in \cite{song2023intelligent} estimated the target angle with respect to the passive IRS based on the echo signals over the BS-IRS-target-IRS-BS link. The work \cite{chen2023ris} considered a bi-static sensing system assisted by passive IRSs, estimating the angle information of perceived objects based on the BS-target-IRS-sensor link. The works \cite{huang2023joint,lin2021channel} studied the user localization and environmental sensing problems in uplink communication systems aided by IRSs in different scenarios. The works \cite{emenonye2023ris,dardari2021nlos} investigated user device localization based on near-field and far-field channel models in downlink communication systems. The works \cite{lu2021target,buzzi2021radar,buzzi2022foundations} considered IRS-assisted target detection and optimized the passive beamforming of IRSs to maximize the target detection probability under false alarm probability constraints. Nevertheless, in these works, transmit signals generally suffer from significant path loss caused by multi-hop reflections, which forms the bottleneck for further improving the system performance \cite{buzzi2022foundations}. 

To overcome this limitation, the {\it active} IRS architecture \cite{zhang2022active} is becoming a new viable solution. In contrast to passive IRSs that only reflect signals without amplification, active IRSs have the capability of amplifying reflecting signals via integrating reflection-type amplifiers into reflecting elements. Despite the additional power consumption, active IRS can effectively compensate for the severe path loss in an energy-efficient manner \cite{kang2024active}. The merits of {active IRSs} make them excellent enablers for both communication and sensing systems. Several works conducted preliminary studies on the application of active IRSs in integrated sensing and communication (ISAC) systems. For instance, in \cite{salem2022active}, the authors proposed the utilization of an active IRS to enhance the communication secrecy rate while ensuring the minimum radar detection SNR. Furthermore, \cite{zhang2022CRAN} investigated the active IRS-aided ISAC in a cloud radio access network, where an active IRS is adopted to address the blockage issue between the BS and targets/users. The radar beampattern towards the sensing targets was optimized to boost the sensing performance. Similarly, the work \cite{zhu2023joint} also deployed an active IRS to introduce an additional virtual LoS link between the BS and the target. The study jointly designed the transmit/receive and reflective beamforming to maximize the radar SNR while ensuring the predefined signal-to-interference-plus-noise ratios (SINRs) for communication users. 

However, the above works on passive and active IRS-enabled sensing or ISAC mainly focused on the case with one single IRS. Unfortunately, a single IRS has a limited coverage area that cannot cover different LoS-blocked areas around the BS due to the round-trip path loss. On the other hand, for sensing tasks such as direction-of-arrival (DoA)-based target localization, a single IRS only provides a piece of sensing information inferred from one observation angle, which is not robust and far from satisfactory for target sensing. The prior work \cite{fang2023multiirsenabled} studied a muti-IRS-enabled ISAC in which multiple passive IRSs are used to extend the coverage areas of both sensing and communications. How to leverage multiple active IRSs for achieving efficient multi-view sensing and improving the performance and robustness of sensing is an uncharted area, which motivates our study in this work.

In this paper, we investigate the multi-active-IRS cooperative sensing system, in which multiple active IRSs are deployed at different locations to facilitate wide-area multi-view sensing and overcome the severe path loss due to multi-hop reflections. As compared to the single-passive-IRS counterpart, the multi-active-IRS cooperative sensing brings new technical challenges due to the {\it active} and {\it multiple} features. First, the active IRS brings additional additive white Gaussian noise (AWGN) in the reflection signal. This thus makes it difficult to characterize the corresponding sensing performance such as Cram\'{e}r-Rao bound (CRB) for parameter estimation. Second, the inter-IRS interference is another issue that limits the sensing performance. Third, how to jointly design transmit beamforming at the BS and reflective beamforming at active IRSs is also a paramount problem demanding prompt solutions. To overcome these issues, we propose a multi-IRS cooperative sensing framework and derive the closed-form CRBs for target's DoA estimation. Based on the derived CRB expression, we propose efficient transmit and reflective beamforming algorithms to improve the sensing performance. The main results of this paper are summarized as follows.
\begin{itemize}
\item First, we propose a multi-IRS cooperative sensing framework, in which the IRSs operate in a time division mode to avoid the inter-IRS interference. In addition, we introduce two different configurations of active IRSs, i.e., with and without dedicated sensors, corresponding to the cases of sensing at the BS and at the IRSs, respectively. Under these two setups, we derive the closed-form CRBs for the estimation of target's DoA with respect to each IRS. The derived close-form CRBs reveal the impact of the reflected noise at the active IRSs on target sensing, which is different from passive IRSs.
\item Then, for both sensing setups, we minimize the CRB for target DoA estimation by jointly optimizing the transmit beamforming at the BS and the reflective beamforming at the IRSs. The formulated problems are high non-convex due to the non-convexity of objective function and transmit power constraints at the IRSs. To handle these challenges, we propose two efficient algorithms based on alternating optimization, successive convex approximation (SCA), and semi-definite relaxation (SDR) to obtain high-quality solutions. 
\item Finally, numerical results verify the effectiveness of our proposed design and the advantages of active IRS-assisted sensing compared to that with passive IRSs. It is shown that the proposed designs outperform various benchmark schemes with transmit beamforming only or reflective beamforming only. It is also unveiled that transmit beamforming at the BS is of greater importance than reflective beamforming at IRSs in minimizing the sensing CRB.
\end{itemize}


{\it Notations:} The circularly symmetric complex Gaussian distribution with mean $\bm{\mu}$ and covariance $\mathbf{A}$ are denoted as $\mathcal{CN}(\bm{\mu},\mathbf{A})$. The notations $(\cdot)^{T}$, $(\cdot)^{*}$, $(\cdot)^{H}$, and $\mathrm{tr}(\cdot)$ denote the transpose, conjugate, conjugate-transpose, and trace operators, respectively. $\mathbf{I}_{L}$ stands for the identity matrix of size $L \times L$ and $\mathbf{e}_{i}$ denotes the $i$-th column of the identity matrix $\mathbf{I}_{4}$. $\Re(\cdot)$ and $\Im(\cdot)$ denote the real and imaginary parts of the argument, respectively. $|\cdot|$ and $\mathrm{arg}\left\{\cdot\right\}$ denote the absolute value and angle of a complex element, respectively. $\mathrm{vec}(\cdot)$ denotes the vectorization operator, $\mathbb{ E}(\cdot)$ denotes the expectation operation, $\mathrm{diag}(\mathbf{x})$ denotes a diagonal matrix with the diagonal entries specified by vector $\mathbf{x}$, and $\mathrm{Diag}(\mathbf{X})$ denotes a diagonal matrix with the diagonal entries specified by the diagonal elements in $\mathbf{X}$. $\mathrm {rank}\left(\mathbf{X}\right)$ denotes the rank value of matrix $\mathbf{X}$ and $[\cdot]_{l,p}$ denotes the $(l,p)$-th element of a matrix. $j$ denotes the imaginary unit. $\otimes$ and $\circ$ denote the Kronecker product and Hadamard product operators, respectively.
\section{System Model}
\vspace{-2mm}
\begin{figure}[htbp]
	\setlength{\abovecaptionskip}{-0pt}
	\setlength{\belowcaptionskip}{-5pt}
	\centering
	\includegraphics[width=0.5\textwidth]{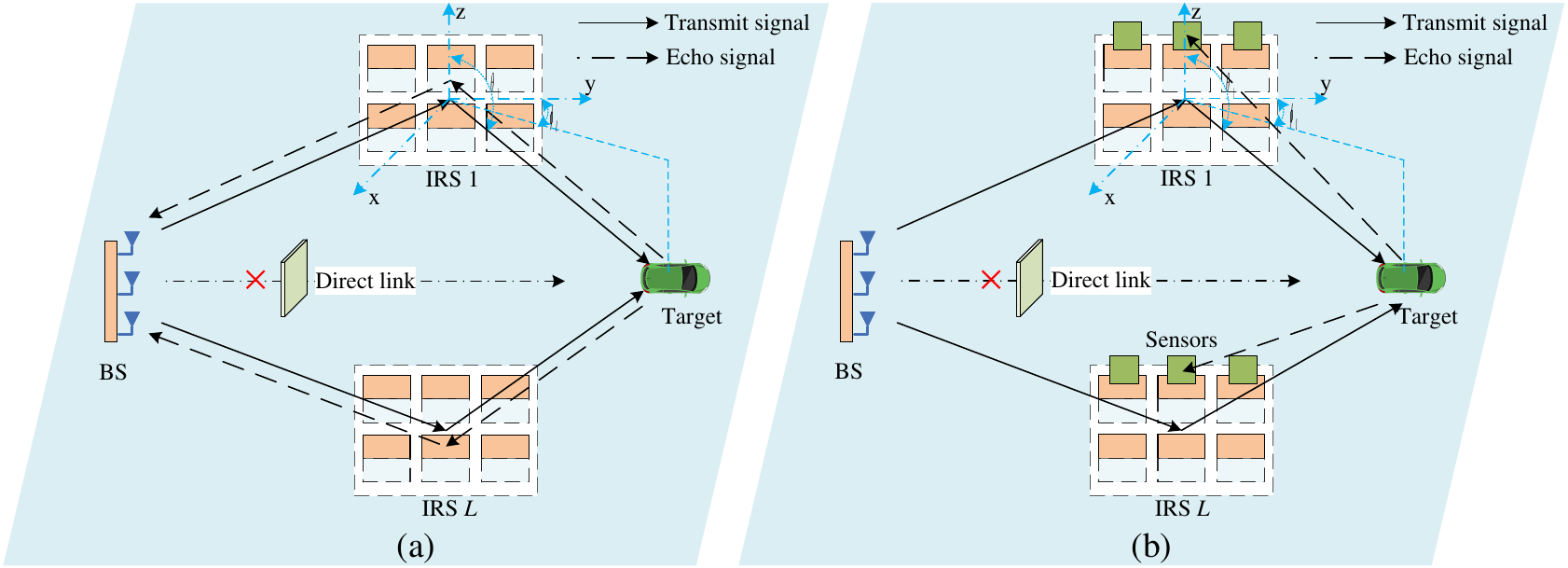}
	\caption{The multi-active-IRS cooperative sensing system, (a): Sensing at the BS; (b): Sensing at the IRSs.}\label{fig:SystemModel}
\end{figure}

We consider a multi-active-IRS cooperative sensing system, where a BS and $L$ active IRSs cooperate to locate one target. The BS is equipped with $M$ uniform linear array (ULA) antennas while each IRS consists of a uniform planar array (UPA) of $N = N_{h}\times N_{v}$ reflecting elements. Let $\mathcal{L} = \{1,\ldots,L\}$ denote the set of IRSs and $\mathcal{N} = \{1,\ldots,N\}$ denote the set of reflecting elements at each IRS in this system. It is assumed that the direct link between the BS and the target is obstructed by blockages such as infrastructures, vehicles, and environmental elements. 

We consider the quasi-static channel model, in which the wireless channels remain unchanged over the transmission block of interest. Let $\mathbf{G}_{l} \in \mathbb{C}^{N \times M}$ denote the channel matrix between the BS and IRS $l$, which can be obtained by the system via proper channel estimation (see, e.g., \cite{zheng2022survey,lin2022channel}). We assume that the $L$ IRSs are delicately deployed such that the channels between IRSs and the target are LoS.\footnote{Note that the $L$ IRSs we consider are selected from all IRSs deployed in the network based on prior target information, such as its approximate location, to ensure a LoS channel between each IRS and the target.} Denote $\theta_{l}$ and $\phi_{l}$ as the vertical and azimuth angles of the target with respect to IRS $l$, respectively. 

Next, we present the model regarding transmit beamforming at the BS and the reflective beamforming at the IRSs. Let $\bm{\psi}_{l} = [\psi_{l,1},\ldots, \psi_{l,N}]^{T}$ denote the complex reflection coefficients imposed by IRS $l$ and $a_{\text{max}}$ denote the maximum amplitude amplification gain of each element at IRS. Since each element of active IRSs can not only tune the phase but also amplify the amplitude of the signal, the complex reflective coefficient is formulated as $\psi_{l,n} = a_{l,n}e^{j\rho_{l,n}}, \forall l\in \mathcal{L}, n\in \mathcal{N}$, where $a_{l,n}$ and $\rho_{l,n}$ are the amplitude and phase of reflective coefficient, respectively \cite{zhang2022active}. The maximum power amplification gain constraints at the active IRSs are expressed as $|\psi_{l,n}|=a_{l,n} \leq a_{\text{max}}, \forall l\in \mathcal{L}, n\in \mathcal{N}$.

In this work, we consider that the IRSs operate in a time division mode to avoid the inter-IRS interference. Let $T_{c}$ denote the number of total symbols in each transmission block of interest or the radar dwell time and $T_L$ denote the duration for each IRS to be active. Here, $T_{c}$ is the multiple of $T_L$, i.e., $T_c = T_L L$. Under this mode, let set $\mathcal{T}_{l} = \{(l-1)\frac{T_{c}}{L}+1,\ldots,l\frac{T_{c}}{L}\}$. Accordingly, at each time symbol $t \in \mathcal T_l$, IRS $l$ is active and the other IRSs are silent. Let $\mathbf{s}_{l}[t] = \left[{s}_{l,1}[t],\dots,s_{l,M}[t]\right]^{T}$ denote the transmitted signal by the BS at time symbol $t\in \mathcal{T}_{l}$. Then, the sample covariance matrix of the transmit signal over the corresponding $T_L$ time symbols is given by
\begin{align}
	\mathbf{R}_{s,l} = \frac{1}{T_{L}} \sum\limits_{t\in \mathcal{T}_{l}}\mathbf{s}_{l}[t]\mathbf{s}_{l}^{H}[t] \succeq 0.
\end{align}
The total sample covariance matrix over the radar dwell time $T_{c}$ needs to satisfy the maximum transmit power constraint $\frac{1}{L}\sum\nolimits_{l \in \mathcal{L}}\mathrm{tr}\left(\mathbf{R}_{s,l}\right) \leq P_{\text{t}}$, where $P_{\text{t}}$ is the maximum transmit power at the BS. In this paper, we consider two sensing strategies that leverage active IRSs with and without dedicated sensors, which are presented in the following two sections. 
\vspace{-1mm}
\section{Sensing at BS: Active IRSs without Dedicated Sensors}
In this section, we consider the sensing at BS as shown in Fig.\ref{fig:SystemModel} (a), with active IRSs without sensors. In this case, all active IRSs only manipulate signals from the BS or echo signals reflected by the target, and the sensing signal reception and processing are implemented at the BS. Recall that $\theta_{l}$ (rad) and $\phi_{l}$ (rad) denote the vertical and azimuth angles at the IRS, respectively. Then, the steering vector of the reflecting elements at IRS $l$ for the DoA $(\theta_{l},\phi_{l})$ is given by
\begin{align}
	\mathbf{a}_{l} = \mathbf{a}_{v}(\theta_l)\otimes\mathbf{a}_{h}(\theta_l,\phi_l),
\end{align}
where
\begin{align}
	 & \mathbf{a}_{v}(\theta_l)\!=\![1,e^{j\frac{2\pi d_v}{\lambda}\cos(\theta_l)},\cdots,e^{j\frac{2\pi d_v(N_v -1)}{\lambda}\cos(\theta_l)}]^{T}\!, \\
	 & \mathbf{a}_{h}(\theta_l,\phi_l)=\nonumber                                                                                                      \\
	 & [1,e^{j\frac{2\pi d_h}{\lambda}\sin(\theta_l)\cos(\phi_l)},\cdots,e^{j\frac{2\pi d_h(N_h -1)}{\lambda}\sin(\theta_l)\cos(\phi_l)}]^{T}\!\!,
\end{align}
and $\lambda$ denotes the carrier wavelength. In addition, $d_h$ and $d_v$ denote the horizontal and vertical spacing of two neighboring reflection elements at IRSs, respectively. Thus, the round-trip target response matrix of the IRS $l$-target-IRS $l$ link is given by
\begin{align}
	\mathbf{E}_{l} = \beta_{l,l}\mathbf{a}_{l}\mathbf{a}_{l}^{T},\label{Target_RM_reflect}
\end{align}
where $\beta_{l,l}$ denotes the complex coefficient that accounts for the radar cross-section of the target and the round-trip path loss of the IRS $l$-target-IRS $l$ link. 

\subsection{Received Echo Signal Model}
In particular, all $L$ IRSs operate in a time division and at each symbol $t \in \mathcal{T}_{l}$, IRS $l$ amplifies the signals twice during one round-trip signal propagation. First, IRS $l$ amplifies the transmit signal $s_{l}[t]$ from the BS as
\begin{align}
	{\mathbf{x}_{l,1}}[t] & =  \mathbf{\Psi}_{l}\mathbf{G}_{l}\mathbf{s}_{l}[t]  + \mathbf{\Psi}_{l}{\mathbf{z}_{l,1}}[t], t \in \mathcal{T}_{l},\label{IRS_ref_sig11}
\end{align}
where $\mathbf{\Psi}_{l} = \text{diag}(\bm{\psi}_{l})$, and ${\mathbf{z}_{l,1}}[t] \sim \mathcal{CN}(\mathbf{0},\sigma_{\text{r}}^2\mathbf{I}_{N})$ denotes the AWGN induced by active IRS $l$. Second, when the target reflects the signal back to IRS $l$, IRS $l$ again amplifies the echo signal as
\begin{align}
	{\mathbf{x}_{l,2}}[t] & = \mathbf{\Psi}_{l}\left(\mathbf{E}_{l}{\mathbf{x}_{l,1}}[t] + {\mathbf{z}_{l,2}}[t]\right)\nonumber                                                                                                                      \\
	                      & =  \mathbf{\Psi}_{l}\mathbf{E}_{l}\mathbf{\Psi}_{l}\mathbf{G}_{l}\mathbf{s}_{l}[t] + \mathbf{\Psi}_{l}\mathbf{E}_{l}\mathbf{\Psi}_{l}{\mathbf{z}_{l,1}}[t] + \mathbf{\Psi}_{l}{\mathbf{z}_{l,2}}[t],\label{IRS_ref_sig12}
\end{align}
where ${\mathbf{z}_{l,2}}[t]\sim \mathcal{CN}(\mathbf{0},\sigma_{\text{r}}^2\mathbf{I}_{N})$ is the AWGN induced by active IRS $l$. Let $P_{\text{s}}$ denote the maximum transmit power budget at each IRS. Therefore, the transmit power constraint at each active IRS is given by
\begin{align}
 & \mathbb{E}\left\{ \|{\mathbf{x}_{l,1}}[t]\|^2 + \|{\mathbf{x}_{l,2}}[t]\|^2 \right\} = \nonumber \\
	        & \mathrm{tr}\left(\mathbf{\Psi}_{l}\mathbf{E}_{l}\mathbf{\Psi}_{l}\mathbf{G}_{l}\mathbf{R}_{s,l}\mathbf{G}_{l}^{H}\mathbf{\Psi}_{l}^{H}\mathbf{E}_{l}^{H}\mathbf{\Psi}_{l}^{H}\right) \nonumber                                                                                                 \\
	        & + \sigma_{\text{r}}^2\mathrm{tr}\left(\mathbf{\Psi}_{l}\mathbf{E}_{l}\mathbf{\Psi}_{l}\mathbf{\Psi}_{l}^{H}\mathbf{E}_{l}^{H}\mathbf{\Psi}_{l}^{H}\right)\!\!+\!\! \mathrm{tr}\left( \mathbf{\Psi}_{l}\mathbf{G}_{l}\mathbf{R}_{s,l}\mathbf{G}_{l}^{H}\mathbf{\Psi}_{l}^{H} \right)  \nonumber \\
	        & + 2 \sigma_{\text{r}}^2 \mathrm{tr}\left(\mathbf{\Psi}_{l}\mathbf{\Psi}_{l}^{H}\right) \leq P_{\text{s}}, \quad \forall l\in \mathcal{L}.\label{IRS_power_cons1}
\end{align}
Based on the signal model at the IRSs in \eqref{IRS_ref_sig12}, the received echo signal at the BS from IRS $l$ at time symbol $t\in \mathcal{T}_{l}$ is given by
\begin{align} \label{ReflectedRadar}
	{\mathbf{y}}_{l}[t] & =  \mathbf{G}_{l}^{T}\mathbf{\Psi}_{l}\mathbf{E}_{l}\mathbf{\Psi}_{l}\mathbf{G}_{l}\mathbf{s}_{l}[t]  + \mathbf{G}_{l}^{T}\mathbf{\Psi}_{l}\mathbf{E}_{l}\mathbf{\Psi}_{l}{\mathbf{z}_{l,1}}[t] \nonumber \\
	                    & + \mathbf{G}_{l}^{T}\mathbf{\Psi}_{l}{\mathbf{z}_{l,2}}[t] + {\mathbf{z}}[t],
\end{align}
where $\mathbf{z}[t]\sim \mathcal{CN}(\mathbf{0},\sigma_{\text{b}}^2\mathbf{I}_{M}) $ denotes the AWGN at the BS.  By concatenating $\mathbf{S}_{l}= [\mathbf{s}_{l}[(l-1)\frac{T_{c}}{L}+1],\ldots,\mathbf{s}_{l}[l\frac{T_{c}}{L}]]$, ${\mathbf{Y}}_{l} = [{\mathbf{y}}_{l}[(l-1)\frac{T_{c}}{L}+1],\ldots,{\mathbf{y}}_{l}[l\frac{T_{c}}{L}]]$, ${\mathbf{Z}}_{l,1} = [{\mathbf{z}}_{l,1}[(l-1)\frac{T_{c}}{L}+1],\ldots,{\mathbf{z}}_{l,1}[l\frac{T_{c}}{L}]]$, and ${\mathbf{Z}}_{l} = [{\mathbf{z}}[(l-1)\frac{T_{c}}{L}+1],\ldots,{\mathbf{z}}[l\frac{T_{c}}{L}]]$, we have
{\begin{align}
	\!\!{\mathbf{Y}}_{l} & \!=\!  \mathbf{G}_{l}^{T}\mathbf{\Psi}_{l}\mathbf{E}_{l}\mathbf{\Psi}_{l}\mathbf{G}_{l}\mathbf{S}_{l}  \!+\! \mathbf{G}_{l}^{T}\mathbf{\Psi}_{l}\mathbf{E}_{l}\mathbf{\Psi}_{l}{\mathbf{Z}_{l,1}} \!\!+\! \mathbf{G}_{l}^{T}\mathbf{\Psi}_{l}{\mathbf{Z}_{l,2}} \!+\! {\mathbf{Z}}_{l}.\label{EchoSigAtBS1}
\end{align}}Accordingly, based on the received echo signal ${\mathbf{Y}}_{l}$ in \eqref{EchoSigAtBS1},  the BS needs to estimate the DoAs $\{\theta_{l},\phi_{l}\}$ and the complex coefficient $\{\beta_{l}\}$ in the complete target response matrix $\mathbf{E}_{l}$ as unknown parameters. Then, based on the estimated DoAs of the target with respect to all IRSs $\{\theta_{l},\phi_{l}\}, l \in \mathcal{L}$, the Stansfield method can be utilized to infer the coordinates of the target \cite{gavish1992performance}.\footnote{After implementing the designed transmit and reflective beamforming in the considered system, the received echo signals by the BS or the sensors at the IRSs are leveraged to estimate the DoA $\{\theta_{l},\phi_{l}\}$ by using sophisticated estimation methods, such as the multiple signal classification (MUSIC) technique \cite{shao2022target}. The design of DoA estimation and target localization algorithms is beyond the scope of this work, and we defer them for future research.}
\vspace{-1mm}
\subsection{Estimation CRB}
For sensing at the BS, the BS needs to estimate the angle of the target regard to the IRSs, i.e., $\{\theta_{l},\phi_{l}\}$. Let ${\bm{\xi}}_{l} = [\theta_{l}, \phi_{l}, {\bm{\beta}_{l}^{T}}]^{T}$ denote the vector of unknown parameters to be estimated with respect to IRS $l$, where ${\bm{\beta}_{l}} = [\Re(\beta_{l}),\Im({\beta_{l}})]^{T}$. By vectorizing the received echo signal $\mathbf{Y}_{l}$ in \eqref{EchoSigAtBS1}, we have
\begin{align}
	{\mathbf{y}}_{l} & = \mathrm{vec}({\mathbf{Y}}_{l})= {\bm{\eta}}_{l} +{\mathbf{w}}_{l},\label{vec_y_l_bs}
\end{align}
where
\begin{align}
	{\bm{\eta}}_{l}  & = \left[(\mathbf{G}_{l}^{T}\mathbf{\Psi}_{l}{\mathbf{E}}_{l}\mathbf{\Psi}_{l}\mathbf{G}_{l}\mathbf{s}_{l}[\tfrac{(l-1)T_{c}}{L}+1])^{T}, \right.\nonumber             \\
	                 & \left.\cdots,(\mathbf{G}_{l}^{T}\mathbf{\Psi}_{l}{\mathbf{E}}_{l}\mathbf{\Psi}_{l}\mathbf{G}_{l}\mathbf{s}_{l}[l\tfrac{T_{c}}{L}])^{T} \right]^{T},\label{vec_sig_bs} \\
	{\mathbf{w}}_{l} & = \left[(\mathbf{G}_{l}^{T}\mathbf{\Psi}_{l}{\mathbf{z}_{l,2}}[\tfrac{(l-1)T_{c}}{L}+1])^{T} + ({\mathbf{z}}_{l}[\tfrac{(l-1)T_{c}}{L}+1])^{T}, \right.\nonumber      \\
	                 & \left.\cdots,(\mathbf{G}_{l}^{T}\mathbf{\Psi}_{l}{\mathbf{z}_{l,2}}[l\tfrac{T_{c}}{l}])^{T} + ({\mathbf{z}}_{l}[l\tfrac{T_{c}}{l}])^{T}\right].\label{vec_noise_bs}
\end{align}
Note that in \eqref{vec_noise_bs}, we ignore the noise term ${\mathbf{z}_{l,1}}$ as it can be relatively weak due to the triple reflection over the IRS-target-IRS-BS link. Based on \eqref{vec_sig_bs} and \eqref{vec_noise_bs}, the mean vector and covariance matrix of ${\mathbf{y}}_{l}$ are obtained as ${\bm{\eta}}_{l}$ and
\begin{align}
	\mathbf{R}_{{\mathbf{y}}_{l}} = \mathbf{I}_{\frac{T_{c}}{L}} \otimes \mathbf{R}_{{\mathbf{w}}_{l}},\label{R_y_bar_bs}
\end{align}
respectively, where $\mathbf{R}_{{\mathbf{w}}_{l}} =\sigma_{\text{r}}^{2}\mathbf{G}_{l}^{T}\mathbf{\Psi}_{l}\mathbf{\Psi}_{l}^{H}\mathbf{G}_{l}^{*}+\sigma_{\text{b}}^{2}\mathbf{I}_{{M}}$. According to the definition of CRB, the CRB for estimating parameter vector ${\bm{\xi}}_{l}$ is given by $\mathrm{CRB}_{{\bm{\xi}}_{l}}(\mathbf{R}_{s,l},\mathbf{\Psi}_{l})=\mathrm{tr}(\mathbf{F}_{l}^{-1})$, where ${\mathbf{F}}_{l} \in \mathbb{R}^{4\times4}$ denotes the Fisher information matrix (FIM) with respect to ${\bm{\xi}}_{l}$. According to the estimation theory, the $(p,q)$-th element of ${\mathbf{F}}_{l}$ is given by \cite{kay1993fundamentals}
\begin{align}
	 & [{\mathbf{F}}_{l}]_{p,q} =\nonumber                                                                                                                                                                                                                                                                                                                                                                                                                                                \\
	 & \mathrm{tr}\left(\mathbf{R}_{{\mathbf{y}}_{l}}^{-1}\frac{\partial \mathbf{R}_{{\mathbf{y}}_{l}}}{\partial [{\bm{\xi}}_{l}]_{p}}\mathbf{R}_{{\mathbf{y}}_{l}}^{-1}\frac{\partial \mathbf{R}_{{\mathbf{y}}_{l}}}{\partial [{\bm{\xi}}_{l}]_{q}}\right) + 2\Re\left(\frac{\partial {\bm{\eta}}_{l}^{H}}{\partial [{\bm{\xi}}_{l}]_{p} } \mathbf{R}_{{\mathbf{y}}_{l}}^{-1} \frac{\partial {\bm{\eta}}_{l}}{\partial [{\bm{\xi}}_{l}]_{q} }\right). \label{FIM_def_bs}
\end{align}

Based on \eqref{FIM_def_bs}, we have the following proposition.
\begin{prop}\label{prop_FIM_bs}
	We define the derivatives of ${{\mathbf{a}}}$ with respect to $\theta_{l}$ and $\phi_{l}$ as $\dot{{\mathbf{a}}}_{\theta_{l}}$ and $\dot{{\mathbf{a}}}_{\phi_{l}}$, respectively. The FIM for estimating ${\mathbf{F}}_{l}$ is given by
	\begin{align}
		{\mathbf{F}}_{l} = \left[\begin{array}{ccc}
				{{F}}_{\theta_{l},\theta_l}                  & {{F}}_{\theta_{l},\phi_{l}}                & {\mathbf{F}}_{\theta_{l},\bm{\beta}_{l}}       \\
				{{F}}_{\theta_{l},\phi_{l}}^{T}              & {{F}}_{\phi_{l},\phi_{l}}                  & {\mathbf{F}}_{\phi_{l},\bm{\beta}_{l}}         \\
				{\mathbf{F}}_{\theta_{l},\bm{\beta}_{l}}^{T} & {\mathbf{F}}_{\phi_{l},\bm{\beta}_{l}}^{T} & {\mathbf{F}}_{{\bm{\beta}_{l}},\bm{\beta}_{l}}
			\end{array}\right],\label{FIM_bs}
	\end{align}
	where
	\begin{subequations}
	\begin{align}
		{{F}}_{\varrho,\varpi}                    & =\frac{{{2 T_c}}}{L}{\left| {{\beta _l}} \right|^2}\mathrm{tr}\left(  \mathbf{C}_{\varrho,l}^H  {\mathbf{R}_{{\mathbf{w}}_{l}}^{ - 1}} \mathbf{C}_{\varpi,l} {\mathbf{R}_{s,l}} \right),\label{F_theta_theta} \\
		{\mathbf{F}}_{\varrho,\bm{\beta}_{l}}       & = \frac{{{2 T_c}}}{L}\Re \left( \beta_{l}\mathrm{tr}\left( \mathbf{C}_{\varrho,l}^H  {\mathbf{R}_{{\mathbf{w}}_{l}}^{ - 1}}\mathbf{H}_{l} {\mathbf{R}_{s,l}} \right)[1, j]\right),\label{F_theta_beta}            \\
		{\mathbf{F}}_{{\bm{\beta}_{l}},\bm{\beta}_{l}} & =  \frac{{{2 T_c}}}{L}\mathrm{tr}\left(  \mathbf{H}_{l}^H   {\mathbf{R}_{{\mathbf{w}}_{l}}^{ - 1}} \mathbf{H}_{l} {\mathbf{R}_{s,l}} \right) \mathbf{I}_{2},\label{F_beta_beta}                                      
	\end{align}
\end{subequations}
	with $\varrho,\varpi \in \{\theta_{l},\phi_{l}\}$, $\mathbf{H}_{l} =\mathbf{G}_{l}^{T}\mathbf{\Psi}_{l} {{\mathbf{a}}}_{l} {\mathbf{a}} _l^T  {{\mathbf{\Psi }}_l}{{\mathbf{G}}_l}$, $\mathbf{C}_{\theta_{l},l} = \mathbf{G}_{l}^{T}\mathbf{\Psi}_{l} \left({\dot{{\mathbf{a}}}_{\theta_{l}} {\mathbf{a}} _l^T + {{ {\mathbf{a}} }_l}{\dot{{\mathbf{a}}}_{\theta_{l}}^T}}\right) {{\mathbf{\Psi }}_l}{{\mathbf{G}}_l}$, and $\mathbf{C}_{\phi_{l},l} = \mathbf{G}_{l}^{T}\mathbf{\Psi}_{l} \left({\dot{{\mathbf{a}}}_{\phi_{l}} {\mathbf{a}} _l^T + {{ {\mathbf{a}} }_l}{\dot{{\mathbf{a}}}_{\phi_{l}}^T}}\right) {{\mathbf{\Psi }}_l}{{\mathbf{G}}_l}$.
\end{prop}
\begin{IEEEproof}
	See Appendix \ref{prop_FIM1_proof}.
\end{IEEEproof}
\begin{rem}
	From \eqref{F_theta_theta}-\eqref{F_beta_beta}, we find that the noise introduced by active IRSs leads to a degradation of the CRB performance. This is because the covariance matrix $\mathbf{R}_{{\mathbf{w}}_{l}}$ in \eqref{R_y_bar_bs} contains an additional term $\sigma_{\text{r}}^{2}\mathbf{G}_{l}^{T}\mathbf{\Psi}_{l}\mathbf{\Psi}_{l}^{H}\mathbf{G}_{l}^{*}$ caused by the noise at the active IRS. Notice that $\mathbf{\Psi}_{l}\mathbf{\Psi}_{l}^{H} = \mathrm{diag}([a_{1,1}^{2},\ldots,a_{l,N}^{2}])$. Therefore, the additional term depends on three factors: the noise variance at the active IRS, the amplitude of the reflective coefficient at the active IRS, and the channel between the BS and the active IRS.
\end{rem}
\subsection{Joint Transmit and Reflective Beamforming Design}
In this subsection, we jointly design the transmit beamforming $\{\mathbf{R}_{s,l}\}$ at the BS and the reflective beamforming $\{\mathbf{\Psi}_{l}\}$ at the IRSs to improve the performance of target estimation for the case with sensing at the BS. Specifically, the optimization problem is formulated by leveraging the closed-form FIM derived in \eqref{FIM_bs}. Accordingly, we propose an algorithm to solve the formulated problem and obtain an efficient solution.

First, our aim is to minimize the maximum CRBs among $\mathrm{CRB}_{\bar{\bm{\xi}}_{l}}(\mathbf{R}_{s,l},\mathbf{\Psi}_{l})$, subject to the constraints on the maximum transmit power at the BS, as well as the maximum transmit power and the maximum power amplification gain at individual IRSs. Consequently, the optimization problem is formulated as 
\begin{subequations}
	\begin{eqnarray}
			\!\!\!(\text{P1}):\!\!\!&{\mathop {\min}\limits_{\{\mathbf{\Psi}_{l},\mathbf{R}_{s,l}\} } } \!\!\!\!&\mathop{\max} \limits_{l \in \mathcal{L}}  \quad \mathrm{CRB}_{\bar{\bm{\xi}}_{l}}(\mathbf{R}_{s,l},\mathbf{\Psi}_{l}) \nonumber\\
		&\text{s.t.} \!\!\!\!\!\!\!& \eqref{IRS_power_cons1},\label{P1_I_cons1}\\
		&& \frac{1}{L}\sum\limits_{l \in \mathcal{L}}\mathrm{tr}\left( \mathbf{R}_{s,l}\right) \leq P_{\text{t}}, \label{P1_I_cons2}\\	
		&&\mathbf{R}_{s,l} \succeq \mathbf{0}, \forall l \in \mathcal{L}, \label{P1_I_cons3}\\
		&&|[\mathbf{\Psi}_{l}]_{n,n}| \leq a_{\text{max}}, \forall l\in \mathcal{L}, n \in \mathcal{N}. \label{P1_I_cons4}
	\end{eqnarray}
\end{subequations}
In problem (P1), \eqref{P1_I_cons1}-\eqref{P1_I_cons4} are the transmit power constraints at the IRSs, the transmit power constraint at the BS, the semi-definite constraint regarding the sample covariance matrix of the transmit signal, and the maximum power amplification gain constraints at the IRSs, respectively. Note that problem (P1) is non-convex due to the non-convexity of the objective function and the constraint in \eqref{P1_I_cons1}. To address this issue, we adopt the alternating optimization approach, wherein the transmit signal covariance $\{\mathbf{R}_{s,l}\}$ at the BS and the reflection coefficients $\{\mathbf{\Psi}_{l}\}$ at the IRSs are optimized alternately.

\subsubsection{Optimal Transmit Signal Covariance under Given $\{\mathbf{\Psi}_{l}\}$}
Under given reflective beamforming $\{\mathbf{\Psi}_{l}\}$, the optimization of transmit beamforming $\{\mathbf{R}_{s,l}\}$ is reformulated as 
	\begin{align}
		     \begin{array}{*{20}{lcl}}
			(\text{P2}):&{\mathop {\min}\limits_{\{\mathbf{R}_{s,l}\} } } &\mathop{\max} \limits_{l \in \mathcal{L}}  \quad \mathrm{tr}(\mathbf{F}_{l}^{-1}) \nonumber\\
			&\text{s.t.} & \eqref{P1_I_cons1}-\eqref{P1_I_cons3}.\nonumber
		\end{array}
	\end{align}
Problem (P2) is also difficult to solve due to the fact that the objective function cannot be expressed in an analytical form. To solve problem (P2), we first introduce an auxiliary variable $\kappa_{i}$ to transform it into the following equivalent problem:
\begin{subequations}
	\begin{eqnarray}
			(\text{P2.1}):&{\mathop {\min}\limits_{ \{\mathbf{R}_{s,l}\},\kappa  } } &  \kappa \nonumber\\
		&\text{s.t.} & \mathrm{tr}(\mathbf{F}_{l}^{-1}) \leq \kappa, \forall l\in \mathcal{L} \label{P2_1_cons4}, \\
		&& \eqref{P1_I_cons1}-\eqref{P1_I_cons3}.\nonumber
	\end{eqnarray}
\end{subequations}
From the expression of FIM in \eqref{FIM_bs}, we note that $\mathbf{F}_{l}$ is a linear function to $\mathbf{R}_{s,l}$ and thus $\mathrm{tr}(\mathbf{F}_{l}^{-1})$ is a convex function to $\mathbf{R}_{s,l}$. Therefore, the constraints in \eqref{P2_1_cons4} are convex. As a result, problem (P2.1) is a convex semi-definite program (SDP), which can be optimally solved by existing solvers like CVX \cite{grant2014cvx}. 

\subsubsection{Reflective Beamforming Optimization under Given $\{\mathbf{R}_{s,l}\}$}
Under given transmit beamforming $\{\mathbf{R}_{s,l}\}$, the optimization of reflective beamforming $\{\mathbf{\Psi}_{l}\}$ is reformulated as
	\begin{align}
			\begin{array}{*{20}{lcl}}
				(\text{P3}):&{\mathop {\min}\limits_{\{\mathbf{\Psi}_{l}\} } } &\mathop{\max} \limits_{l \in \mathcal{L}}  \quad \mathrm{tr}(\mathbf{F}_{l}^{-1}) \nonumber\\
				&\text{s.t.} & \eqref{P1_I_cons1}, \eqref{P1_I_cons4}.\nonumber
			\end{array}
	\end{align}
Note that the FIM $\mathbf{F}_{l}$ depends only on the reflective beamforming $\mathbf{\Psi}_{l}$ at IRS $l$. Thus, problem (P3) can be equivalently decomposed into $L$ subproblems each given by
	\begin{align}
			\begin{array}{*{20}{lcl}}
				(\text{P3.$l$.1}):&{\mathop {\min}\limits_{\mathbf{\Psi}_{l} } } & \mathrm{tr}(\mathbf{F}_{l}^{-1}) \nonumber\\
				&\text{s.t.} & \eqref{P1_I_cons1}, \eqref{P1_I_cons4}.\nonumber
			\end{array}
	\end{align}
	
 This problem is highly non-convex because the transmit power constraints at IRSs in \eqref{P1_I_cons1} and the elements in FIM are non-convex functions with respect to $\{\mathbf{\Psi}_{l}\}$. By introducing auxiliary variables $\{\kappa_i\}_{i=1}^{4}$, problem (\text{P3.$l$.1}) is equivalent to 
 \begin{subequations}
	\begin{eqnarray}
				\!\!\!\!\!(\text{P3.$l$.2}):&\!\!\!\!\!\!\!{\mathop {\min}\limits_{\mathbf{\Psi}_{l},\{\kappa_i\}_{i=1}^{4}  } } & \sum\limits_{i=1}^{4} \kappa_i \nonumber\\
			&\text{s.t.} & \left[\begin{array}{cc}
				\mathbf{F}_{l} & \mathbf{e}_{i} \\
				 \mathbf{e}_{i}^{T}  & \kappa_i 
			\end{array}\right] \succeq 0, i \in \! \{1, \ldots, 4\}, \label{Schur_Com}\\
			&&\eqref{P1_I_cons1}, \eqref{P1_I_cons4},\nonumber
	\end{eqnarray}
\end{subequations}
where constraint \eqref{Schur_Com} is derived by using the Schur complement. Note that problem (P3.$l$.2) is still non-convex due to the constraints in \eqref{P1_I_cons1} and \eqref{Schur_Com}. To handle it, we resort to the SDR and SCA techniques to transform these constraints into convex forms.

With the definition of $\mathbf{\Theta}_{l} = \bm{\psi}_{l}\bm{\psi}_{l}^{H}$, let $\{\mathbf{\Theta}_{l}^{(i)}\}$ be the local point of $\{\mathbf{\Theta}_{l}\}$ at the $i$-th iteration of SCA, and we approximate the objective function $\mathrm{tr}(\mathbf{F}_{l}^{-1})$ according to Lemma \ref{FIM_appr_lem}. 
\begin{lem}\label{FIM_appr_lem}
With a given local point $\mathbf{\Theta}_{l}^{(i)}$, the FIM in \eqref{FIM_bs} is approximated as 
\vspace{-1mm}
\begin{align}
	\hat{{\mathbf{F}}}_{l} = \left[\begin{array}{ccc}
		\hat{{{F}}}_{\theta_{l},\theta_l}&\hat{{{F}}}_{\theta_{l},\phi_{l}}&\hat{{\mathbf{F}}}_{\theta_{l},\bm{\beta}_{l}}\\
		\hat{{{F}}}_{\theta_{l},\phi_{l}}^{T}&\hat{{{F}}}_{\phi_{l},\phi_{l}}&\hat{{\mathbf{F}}}_{\phi_{l},\bm{\beta}_{l}}\\
		\hat{{\mathbf{F}}}_{\theta_{l},\bm{\beta}_{l}}^{T}& \hat{{\mathbf{F}}}_{\phi_{l},\bm{\beta}_{l}}^{T}&\hat{{\mathbf{F}}}_{{\bm{\beta}_{l}},\bm{\beta}_{l}}
	\end{array}\right],\label{FIM_bs_trans}
\end{align}
in which 
\vspace{-1mm}
{\small
\begin{subequations}
	\begin{align}
	\hat{{F}}_{\varrho,\varpi} &= \frac{{{2 T_c}}}{L}{\left| {{\beta _l}} \right|^2} \mathrm{tr}\left(\nabla_{\varrho,\varpi}^{T}(\mathbf{\Theta}_{l}^{(i)}) \left(\mathbf{\Theta}_{l}-\mathbf{\Theta}_{l}^{(i)}\right)  \right) \nonumber\\
	&+ \frac{{{2 T_c}}}{L}{\left| {{\beta _l}} \right|^2} Q_{\varrho,\varpi}(\mathbf{\Theta}_{l}^{(i)}),\label{hat_F_theta_theta}\\
   \hat{\mathbf{F}}_{\varrho,\bm{\beta}_{l}} &= \frac{{{2 T_c}}}{L}\Re \left( \beta_{l} \left( \mathrm{tr} \left(\nabla_{\varrho,\bm{\beta}_{l}}^{T}(\mathbf{\Theta}_{l}^{(i)}) \left(\mathbf{\Theta}_{l} - \mathbf{\Theta}_{l}^{(i)}\right) \right) \right.\right. \nonumber\\
   &\left.\left. + Q_{\varrho,\bm{\beta}_{l}}(\mathbf{\Theta}_{l}^{(i)}) \right) [1, j]\right),\label{hat_F_phi_beta}\\
   \hat{\mathbf{F}}_{{\bm{\beta}_{l}},\bm{\beta}_{l}} &=  \frac{{{2 T_c}}}{L} \left( \mathrm{tr}\left(\nabla_{{\bm{\beta}_{l}},\bm{\beta}_{l}}^{T}(\mathbf{\Theta}_{l}^{(i)})\left(\mathbf{\Theta}_{l} - \mathbf{\Theta}_{l}^{{i}}\right)  \right) \right. \nonumber\\
   &\left.+ Q_{{\bm{\beta}_{l}},\bm{\beta}_{l}}(\mathbf{\Theta}_{l}^{(i)}) \right) \mathbf{I}_{2}.\label{hat_F_beta_beta}
\end{align}
\end{subequations}}
\end{lem}
\begin{IEEEproof}
	See Appendix \ref{FIM_appr_proof}.
\end{IEEEproof}

Then, we transform the constraint in \eqref{P1_I_cons1} into a convex form. The four terms in the left-hand-side of \eqref{P1_I_cons1} are equivalent to 
\vspace{-1mm}
\begin{subequations} 
	\begin{align}
	&\mathrm{tr}\left(\mathbf{\Psi}_{l}\mathbf{E}_{l}\mathbf{\Psi}_{l}\mathbf{G}_{l}\mathbf{R}_{s,l}\mathbf{G}_{l}^{H}\mathbf{\Psi}_{l}^{H}\mathbf{E}_{l}^{H}\mathbf{\Psi}_{l}^{H}\right)\nonumber\\
	&= \left| {{\beta _l}} \right|^2 \mathrm{tr}\left(\mathbf{\Psi}_{l} \mathbf{a}_{l}\mathbf{a}_{l}^{T} \mathbf{\Psi}_{l}\mathbf{G}_{l}\mathbf{R}_{s,l}\mathbf{G}_{l}^{H}\mathbf{\Psi}_{l}^{H}\mathbf{a}_{l}^{*} \mathbf{a}_{l}^{H} \mathbf{\Psi}_{l}^{H}\right) \nonumber\\
	&= \left| {{\beta _l}} \right|^2 \mathrm{tr}\left(\mathbf{A}_{l}^{H}\mathbf{A}_{l}\mathbf{\Theta}_{l}\right) \mathrm{tr}\left(\mathbf{R}_{1}\mathbf{\Theta}_{l}^{T} \right), \label{P1_I_cons1_1}\\
	& \mathrm{tr}\left(\mathbf{\Psi}_{l}\mathbf{E}_{l}\mathbf{\Psi}_{l}\mathbf{\Psi}_{l}^{H}\mathbf{E}_{l}^{H}\mathbf{\Psi}_{l}^{H}\right) \nonumber\\
	&= \left| {{\beta _l}} \right|^2 \mathrm{tr}\left(\mathbf{\Psi}_{l}\mathbf{a}_{l}\mathbf{a}_{l}^{T}\mathbf{\Psi}_{l}\mathbf{\Psi}_{l}^{H}\mathbf{a}_{l}^{*}\mathbf{a}_{l}^{H}\mathbf{\Psi}_{l}^{H}\right) \nonumber\\
	&= \left| {{\beta _l}} \right|^2 \left(\mathrm{tr}\left(\mathbf{A}_{l}^{H}\mathbf{A}_{l}\mathbf{\Theta}_{l}\right)\right)^{2}, \label{P1_I_cons1_2}\\
	& \mathrm{tr}\left( \mathbf{\Psi}_{l}\mathbf{G}_{l}\mathbf{R}_{s,l}\mathbf{G}_{l}^{H}\mathbf{\Psi}_{l}^{H} \right) = \mathrm{tr}\left(\mathbf{G}_{l}\mathbf{R}_{s,l}\mathbf{G}_{l}^{H}\mathrm{Diag}\left(\mathbf{\Theta}_{l}\right) \right), 	 \label{P1_I_cons1_3}\\  
	&\mathrm{tr}\left(\mathbf{\Psi}_{l}\mathbf{\Psi}_{l}^{H}\right) = \mathrm{tr}\left(\mathbf{\Theta}_{l}\right), \label{P1_I_cons1_4}
\end{align}
\end{subequations}
respectively, where $\mathbf{A}_{l} = \mathrm{diag}(\mathbf{a}_{l})$ and $\mathbf{R}_{1} = \mathbf{A}_{l}{{\mathbf{G}}_l} {\mathbf{R}_{s,l}} \mathbf{G}_{l}^{H} \mathbf{A}_{l}^{H}$. Note that \eqref{P1_I_cons1_1} and \eqref{P1_I_cons1_2} are quadratic functions with respect to $\mathbf{\Theta}_{l}$, and we approximate them using their first-order Taylor expansions. The derivatives of the trace terms in \eqref{P1_I_cons1_1} and \eqref{P1_I_cons1_2} with respect to $\mathbf{\Theta}_{l}$ are given by
\begin{subequations} 
\begin{align}
	\nabla_{\text{c},1}(\mathbf{\Theta}_{l}) & = \frac{\partial}{\partial \mathbf{\Theta}_{l}}   \mathrm{tr}\left(\mathbf{A}_{l}^{H}\mathbf{A}_{l}\mathbf{\Theta}_{l}\right) \mathrm{tr}\left(\mathbf{R}_{1}\mathbf{\Theta}_{l}^{T} \right) \nonumber\\
	&= \mathrm{tr}\left(\mathbf{R}_{1}\mathbf{\Theta}_{l}^{T} \right)\mathbf{A}_{l}^{T}\mathbf{A}_{l}^{*} + \mathrm{tr}\left(\mathbf{A}_{l}^{H}\mathbf{A}_{l}\mathbf{\Theta}_{l}\right)\mathbf{R}_{1}, \label{P1_I_cons1_1_der}\\
	\nabla_{\text{c},2}(\mathbf{\Theta}_{l}) & = \frac{\partial}{\partial \mathbf{\Theta}_{l}}  \left(\mathrm{tr}\left(\mathbf{A}_{l}^{H}\mathbf{A}_{l}\mathbf{\Theta}_{l}\right)\right)^{2} \nonumber\\
	&= 2\mathrm{tr}\left(\mathbf{A}_{l}^{H}\mathbf{A}_{l}\mathbf{\Theta}_{l}\right)\mathbf{A}_{l}^{T}\mathbf{A}_{l}^{*}. \label{P1_I_cons1_2_der}
\end{align}
\end{subequations}
Based on \eqref{P1_I_cons1_3} and \eqref{P1_I_cons1_4}, and the derivatives in \eqref{P1_I_cons1_1_der} and \eqref{P1_I_cons1_2_der}, the constraint in \eqref{P1_I_cons1} is approximated as
\begin{align}
	&\!\! \!\! \left| {{\beta _l}} \right|^2  \mathrm{tr}\left(\nabla_{\text{c},1}^{T}(\mathbf{\Theta}_{l}^{i})\left(\mathbf{\Theta}_{l}-\mathbf{\Theta}_{l}^{(i)}\right)\right)  \nonumber\\
	&\!\! \!\! + \left| {{\beta _l}} \right|^2 \mathrm{tr}\left(\mathbf{A}_{l}^{H}\mathbf{A}_{l}\mathbf{\Theta}_{l}^{(i)}\right) \mathrm{tr}\left(\mathbf{R}_{1}\left(\mathbf{\Theta}_{l}^{(i)}\right)^{T} \right)  \nonumber\\
	&\!\! \!\!  + \sigma_{\text{r}}^2 \left| {{\beta _l}} \right|^2  \mathrm{tr}\left(\nabla_{\text{c},2}^{T}(\mathbf{\Theta}_{l}^{i})\left(\mathbf{\Theta}_{l}-\mathbf{\Theta}_{l}^{(i)}\right)\right) \nonumber\\
	&\!\! \!\! + \sigma_{\text{r}}^2 \left| {{\beta _l}} \right|^2 \left|\mathrm{tr}\left(\mathbf{A}_{l}^{H}\mathbf{A}_{l}\mathbf{\Theta}_{l}^{(i)}\right)\right|^{2} \nonumber\\
	&\!\! \!\! +\! \mathrm{tr}\left(\mathbf{G}_{l}\mathbf{R}_{s,l}\mathbf{G}_{l}^{H}\mathrm{Diag}\left(\mathbf{\Theta}_{l}\right) \right) \!+\! 2 \sigma_{\text{r}}^2 \mathrm{tr}\left(\mathbf{\Theta}_{l}\right) \leq P_{\text{s}}, \forall l\in \mathcal{L}. \label{P1_I_cons1_trans}
\end{align}

By adopting \eqref{FIM_bs_trans} and \eqref{P1_I_cons1_trans}, in the $i$-th SCA iteration, problem (P3.$l$.2) is transformed into problem (P3.$l$.3.$i$) with 
\begin{subequations}
	\begin{eqnarray}
				\!\!\!\!\!(\text{P3.$l$.3.$i$}):\!\!\!\!\!\!\!\!&{\mathop {\min}\limits_{\mathbf{\Theta}_{l},\{\kappa_i\}_{i=1}^{4}  } } \!\!\!\!& \sum\nolimits_{i=1}^{4} \kappa_i \nonumber\\
			&\text{s.t.} & \left[\begin{array}{cc}
				\hat{\mathbf{F}}_{l} & \mathbf{e}_{i} \\
				 \mathbf{e}_{i}^{T}  & \kappa_i 
			\end{array}\right] \succeq 0,  i \in \! \{1, \ldots, 4\}, \label{Schur_Com_trans}\\
			&& [\mathbf{\Theta}_{l}]_{n,n} \leq a_{\text{max}}^{2}, \label{amplify_cons}\\
			&& \mathrm{rank}(\mathbf{\Theta}_{l} ) = 1, \label{rank1_cons}\\
			&&\eqref{P1_I_cons1_trans}.\nonumber
	\end{eqnarray}
\end{subequations}
Furthermore, we drop the rank-one constraint in \eqref{rank1_cons} and accordingly obtain the relaxed version of (P3.$l$.3.$i$) as (SDR3.$l$.3.$i$). Note that problem (SDR3.$l$.3.$i$) is a convex problem which can be efficiently solved by CVX. Let $\mathbf{\Theta}_{l}^{\star(i)}$ denote the obtained solution to problem (SDR3.$l$.3.$i$) at the $i$-th SCA iteration, which is then updated as $\mathbf{\Theta}_{l}^{\star(i+1)}$. Note that the objective value achieved by $\mathbf{\Theta}_{l}^{\star(i+1)}$ is always no greater than that by $\mathbf{\Theta}_{l}^{\star(i)}$. Thus, the achieved CRB is monotonically non-increasing after each iteration of SCA. Moreover, the optimal value of problem (P3.$l$.2) is lower bounded due to the non-negativity of the entries of FIM. Based on the above two observations, the convergence of the SCA is guaranteed and we have $\mathbf{\Theta}_{l}^{\star}$ as the corresponding converged solution. Note that  $\mathbf{\Theta}_{l}^{\star}$ may not meet the rank-one condition. Therefore, we implement the Gaussian randomization to find an efficient rank-one solution to (P3.$l$.2) based on the obtained $\{\mathbf{\Theta}_{l}^{\star}\}$ \cite{luo2010semidefinite}. In particular, we generate a number of random vectors ${\mathbf{r}}_{l} \sim \mathcal{CN}\left(\mathbf{0},\mathbf{I}_N\right)$, and then construct a number of rank-one solutions as 
\begin{align}
\boldsymbol{\psi}_{l} = \left(\boldsymbol{\Theta}_{l}^{\star}\right)^{\frac{1}{2}}\mathbf{r}_{l}.
\end{align} 
Next, we verify whether the maximum amplitude of the elements in $\boldsymbol{\psi}_{l}$ exceeds the maximum amplification gain $a_{\text{max}}$. If so, we normalize $\boldsymbol{\psi}_{l} = a_{\text{max}}\frac{\boldsymbol{\psi}_{l}}{\max(|\boldsymbol{\psi}_{l}|)}$, where $\max(|\boldsymbol{\psi}_{l}|)$ represents the maximum amplitude value of the elements in $\boldsymbol{\psi}_{l}$. Finally, we seek the optimal solution of $\boldsymbol{\psi}_{l}$ that minimizes $\mathrm{CRB}_{\bar{\bm{\xi}}_{l}}(\mathbf{R}_{s,l},\mathbf{\Psi}_{l})$ while satisfying the constraints in \eqref{P1_I_cons1} among all randomly generated $\boldsymbol{\psi}_{l}$'s. 

Through alternately solving problems (P2) and (P3), a high-quality solution to problem (P1) is obtained. Note that problem (P2) is optimally solved, while solving (P3) with a sufficiently large number of Gaussian randomizations leads to a non-increasing sequence of max-CRB values. Consequently, the alternating optimization-based algorithm has guaranteed convergence by generating a monotonically decreasing sequence of max-CRB values throughout the iterations. 

\vspace{-1mm}
\section{Sensing at IRSs: Active IRSs with Dedicated Sensors}
In this section, we consider sensing at IRSs as shown in Fig.\ref{fig:SystemModel} (b), with active IRSs with dedicated sensors, and the sensing signal reception and processing are operated at the IRSs. In this case, we assume that each IRS is equipped with a UPA of $\bar{N} = \bar{N}_{h}\times\bar{N}_{v}$ sensors. The steering vector of the sensor elements at IRS $l$ for DoA $(\theta_{l},\phi_{l})$ is given by
\begin{align}
	\bar{\mathbf{a}}_{l} = \bar{\mathbf{a}}_{v}(\theta_l)\otimes\bar{\mathbf{a}}_{h}(\theta_l,\phi_l),
\end{align}
where
\begin{align}
	 & \bar{\mathbf{a}}_{v}(\theta_l)\!=\![1,e^{j\frac{2\pi \bar{d}_v}{\lambda}\cos(\theta_l)},\cdots,e^{j\frac{2\pi \bar{d}_v(\bar{N}_v -1)}{\lambda}\cos(\theta_l)}]^{T}\!, \\
	 & \bar{\mathbf{a}}_{h}(\theta_l,\phi_l)=\nonumber                                                                                                                        \\
	 & [1,e^{j\frac{2\pi \bar{d}_h}{\lambda}\sin(\theta_l)\cos(\phi_l)},\cdots,e^{j\frac{2\pi \bar{d}_h(\bar{N}_h -1)}{\lambda}\sin(\theta_l)\cos(\phi_l)}]^{T}\!\!,
\end{align}
$\bar{d}_h$ and $\bar{d}_v$ denote the horizontal and vertical spacing of two neighboring sensor elements at IRSs, respectively. Thus, the round-trip target response matrix of the IRS $l$-target-IRS $l$ link is given by
\begin{align}
	\bar{\mathbf{E}}_{l} = \beta_{l,l}\bar{\mathbf{a}}_{l}\mathbf{a}_{l}^{T}.\label{Target_RM_sensor}
\end{align}
\subsection{Received Echo Signal Model}
Then, we consider that all IRSs are installed with dedicated sensors for signal reception and the sensing signals are processed at the IRSs. We also assume that the channel state information of $\mathbf{G}_{l}$ is available at IRS $l$ via proper channel estimation. We consider that the  $L$ IRSs operate in a time division manner. At each symbol $t \in \mathcal{T}_{l}$, only IRS $l$ is active and the other IRSs are silent. In this case, each IRS  reflects the signal only once, and the reflected signal at IRS $l$ is given in \eqref{IRS_ref_sig11}. As such, the received echo signal ${\bar{\mathbf{y}}_{l}}[t] \in \mathbb{C}^{\bar{N}\times 1}$ by IRS $l$ at time symbol $t\in \mathcal{T}_{l}$ is given by
\begin{align}
	{\bar{\mathbf{y}}_{l}}[t] = \bar{\mathbf{E}}_{l}\mathbf{\Psi}_{l}\mathbf{G}_{l}\mathbf{s}_{l}[t]  + \bar{\mathbf{E}}_{l}\mathbf{\Psi}_{l}{\mathbf{z}_{l,1}}[t]  + \bar{\mathbf{z}}_{l}[t],
\end{align}
where $\bar{\mathbf{z}}_{l}[t]\sim \mathcal{CN}(\mathbf{0},\sigma_{\text{s}}^2\mathbf{I}_{\bar{N}}) $ denotes the Gaussian noise at the sensor receiver of IRS $l$. By defining $\bar{\mathbf{Y}}_{l} = [\bar{\mathbf{y}}_{l}[(l-1)\frac{T_{c}}{L}+1],\ldots,\bar{\mathbf{y}}_{l}[l\frac{T_{c}}{L}]]$, and $\bar{\mathbf{Z}}_{l} = [\bar{\mathbf{z}}_{l}[(l-1)\frac{T_{c}}{L}+1],\ldots,\bar{\mathbf{z}}_{l}[l\frac{T_{c}}{L}]]$, we have
\begin{align}
	\bar{\mathbf{Y}}_{l} & =  \bar{\mathbf{E}}_{l}\mathbf{\Psi}_{l}\mathbf{G}_{l}\mathbf{S}_{l}  + \bar{\mathbf{E}}_{l}\mathbf{\Psi}_{l}{\mathbf{Z}_{l,1}} + \bar{\mathbf{Z}}_{l},\label{EchoSigAtIRS1}
\end{align}
and each active IRS must satisfy the following transmit power constraints:
\begin{align}
	\mathrm{tr}\left( \mathbf{\Psi}_{l}\mathbf{G}_{l}\mathbf{R}_{s,l}\mathbf{G}_{l}^{H}\mathbf{\Psi}_{l}^{H} \right) +  \sigma_{\text{r}}^2 \mathrm{tr}\left(\mathbf{\Psi}_{l}\mathbf{\Psi}_{l}^{H}\right) \leq P_{\text{s}}, \forall l\in \mathcal{L}. \label{IRS_PowerCons12}
\end{align}

Note that similar as in \eqref{EchoSigAtBS1}, the active IRSs bring the additional AWGN terms in \eqref{EchoSigAtIRS1}, which affect the estimation CRB of target sensing. In the following, we will derive the estimation CRB for characterizing the performance of sensing.

\subsection{Estimation CRB}
Each IRS $l$ needs to estimate the relative direction of the target, i.e., the DoA information $(\theta_{l},\phi_{l})$. The vector of unknown parameters to be estimated at IRS $l$ is ${\bm{\xi}}_{l} = [\theta_{l}, \phi_{l}, {\bm{\beta}_{l}^{T}}]^{T}$. By vectorizing the received echo signal $\bar{\mathbf{Y}}_{l}$ in \eqref{EchoSigAtIRS1}, we have
\begin{align}
	\bar{\mathbf{y}}_{l} & = \mathrm{vec}(\bar{\mathbf{Y}}_{l})= \bar{\bm{\eta}}_{l} + \bar{\mathbf{w}}_{l},\label{vec_y_l_irs}
\end{align}
where
\begin{align}
	 & \bar{\bm{\eta}}_{l} = \left[(\bar{\mathbf{E}}_{l}\mathbf{\Psi}_{l}\mathbf{G}_{l}\mathbf{s}_{l}[\tfrac{(l-1)T_{c}}{L}+1])^{T},\right. \nonumber                                             \\
	 & \left.\cdots,(\bar{\mathbf{E}}_{l}\mathbf{\Psi}_{l}\mathbf{G}_{l}\mathbf{s}_{l}[l\tfrac{T_{c}}{L}])^{T} \right]^{T},\label{vec_sig_irs}                                                    \\
	 & \bar{\mathbf{w}}_{l} = \left[(\bar{\mathbf{E}}_{l}\mathbf{\Psi}_{l}{\mathbf{z}_{l,1}}[\tfrac{(l-1)T_{c}}{L}+1])^{T} + (\bar{\mathbf{z}}_{l}[\tfrac{(l-1)T_{c}}{L}+1])^{T},\right.\nonumber \\
	 & \left.\cdots,(\bar{\mathbf{E}}_{l}\mathbf{\Psi}_{l}{\mathbf{z}_{l,1}}[l\tfrac{T_{c}}{l}])^{T} + (\bar{\mathbf{z}}_{l}[l\tfrac{T_{c}}{l}])^{T}\right].\label{vec_noise_irs}
\end{align}
Based on \eqref{vec_sig_irs} and \eqref{vec_noise_irs}, the mean vector and covariance matrix of $\bar{\mathbf{y}}_{l}$ are obtained as $\bar{\bm{\eta}}_{l}$ and
\begin{align}
	\mathbf{R}_{\bar{\mathbf{y}}_{l}} = \mathbf{I}_{\frac{T_{c}}{L}} \otimes \mathbf{R}_{\bar{\mathbf{w}}_{l}},\label{R_y_bar}
\end{align}
respectively, where $\mathbf{R}_{\bar{\mathbf{w}}_{l}} = \sigma_{\text{r}}^{2}\bar{\mathbf{E}}_{l}\mathbf{\Psi}_{l}\mathbf{\Psi}_{l}^{H}\bar{\mathbf{E}}_{l}^{H}+\sigma_{\text{s}}^{2}\mathbf{I}_{\bar{N}}$. Then, the CRB for estimating parameter vector ${\bm{\xi}}_{l}$ is given by $\overline{\mathrm{CRB}}_{{\bm{\xi}}_{l}}(\mathbf{R}_{s,l},\mathbf{\Psi}_{l})=\mathrm{tr}(\bar{\mathbf{F}}_{l}^{-1})$, where $\bar{\mathbf{F}}_{l} \in \mathbb{R}^{4\times4}$ is the FIM with respect to ${\bm{\xi}}_{l}$. Similar to \eqref{FIM_def_bs}, the $(p,q)$-th element of $\bar{\mathbf{F}}_{l}$ is given by
\begin{align}
	 & [\bar{\mathbf{F}}_{l}]_{p,q} = \nonumber                                                                                                                                                                                                                                                                                                                                                                                                                                                                    \\
	 & \mathrm{tr}\left(\mathbf{R}_{\bar{\mathbf{y}}_{l}}^{-1}\frac{\partial \mathbf{R}_{\bar{\mathbf{y}}_{l}}}{\partial [{\bm{\xi}}_{l}]_{p}}\mathbf{R}_{\bar{\mathbf{y}}_{l}}^{-1}\frac{\partial \mathbf{R}_{\bar{\mathbf{y}}_{l}}}{\partial [{\bm{\xi}}_{l}]_{q}}\right) + 2\Re\left(\frac{\partial \bar{\bm{\eta}}_{l}^{H}}{\partial [{\bm{\xi}}_{l}]_{p} } \mathbf{R}_{\bar{\mathbf{y}}_{l}}^{-1} \frac{\partial \bar{\bm{\eta}}_{l}}{\partial [{\bm{\xi}}_{l}]_{q} }\right). \label{FIM_def}
\end{align}
We define $\mathbf{\Theta}_{l} = \bm{\psi}_{l}\bm{\psi}_{l}^{H}$, where $\mathrm{rank}(\mathbf{\Theta}_{l} ) = 1$. Then, based on \eqref{FIM_def}, we have the following proposition.
\begin{prop}\label{prop_FIM2}
	We define the derivatives of ${\bar{\mathbf{a}}}$ with respect to $\theta_{l}$ and $\phi_{l}$ as $\dot{\bar{\mathbf{a}}}_{\theta_{l}}$ and $\dot{\bar{\mathbf{a}}}_{\phi_{l}}$, respectively. The FIM for estimating $\bar{\mathbf{F}}_{l}$ is given by
	\begin{align}
		\bar{\mathbf{F}}_{l} = \left[\begin{array}{ccc}
				\bar{{F}}_{\theta_{l},\theta_l}                  & \bar{{F}}_{\theta_{l},\phi_{l}}                & \bar{\mathbf{F}}_{\theta_{l},\bm{\beta}_{l}}       \\
				\bar{{F}}_{\theta_{l},\phi_{l}}^{T}              & \bar{{F}}_{\phi_{l},\phi_{l}}                  & \bar{\mathbf{F}}_{\phi_{l},\bm{\beta}_{l}}         \\
				\bar{\mathbf{F}}_{\theta_{l},\bm{\beta}_{l}}^{T} & \bar{\mathbf{F}}_{\phi_{l},\bm{\beta}_{l}}^{T} & \bar{\mathbf{F}}_{{\bm{\beta}_{l}},\bm{\beta}_{l}}
			\end{array}\right],\label{FIM_irs}
	\end{align}
	where
	\begin{subequations}
	{\small\begin{align}
		 & \bar{{F}}_{\varrho,\varpi} = {\frac{4T_{c}}{L}}\sigma _r^4{\left| {{\beta _l}} \right|^4} \left(\mathrm{tr}\left(\mathbf{\Theta}_{l}\right)\right)^{2}  {\mathrm{tr}}\left( {\mathbf{R}_{\bar{\mathbf{w}}_{l}}^{ - 1}}\bar{\mathbf{B}}_{\varrho,l}{\mathbf{R}_{\bar{\mathbf{w}}_{l}}^{ - 1}}\bar{\mathbf{B}}_{\varpi,l} \right) \nonumber             \\
		 & + \frac{{{2 T_c}}}{L} {\left| {{\beta _l}} \right|^2}\mathrm{tr}\left(  {\bar{\mathbf{C}}_{\varrho,l}^H}{\mathbf{R}_{\bar{\mathbf{w}}_{l}}^{ - 1}} \bar{\mathbf{C}}_{\varpi,l} {\mathbf{R}_{s,l}}  \right), \label{F_bar_theta_theta} \\
		 & \bar{\mathbf{F}}_{\varrho,\bm{\beta}_{l}}= \frac{{4{T_c}}}{L}\sigma _r^4{\left| {{\beta _l}} \right|^2} \left(\mathrm{tr}\left(\mathbf{\Theta}_{l}\right)\right)^{2}  {\mathrm{tr}}\left({\mathbf{R}_{\bar{\mathbf{w}}_{l}}^{ - 1}}\bar{\mathbf{B}}_{\varrho,l}{\mathbf{R}_{\bar{\mathbf{w}}_{l}}^{ - 1}}  {\bar{\mathbf{D}}_{l,l}} \right)\times \nonumber \\
		 & [\Re \left( {{\beta _l}} \right),\Im \left( {{\beta _l}} \right)] \!+\! \frac{{{2 T_c}}}{L}\Re \left( \beta_{l}\mathrm{tr}\left({ \bar{\mathbf{C}}_{\varrho,l}^H}{\mathbf{R}_{\bar{\mathbf{w}}_{l}}^{ - 1}}\bar{\mathbf{H}}_{l}{\mathbf{R}_{s,l}}  \right)[1, j]\right),\label{F_bar_theta_beta}                                                                     \\
		 & \bar{\mathbf{F}}_{{\bm{\beta}_{l}},\bm{\beta}_{l}} = \! \frac{{{4T_c}}}{L} \sigma_r^4 \left(\mathrm{tr}\left(\mathbf{\Theta}_{l}\right)\right)^{2}  {\mathrm{tr}}\left( {\mathbf{R}_{\bar{\mathbf{w}}_{l}}^{ - 1}} {\bar{\mathbf{D}}_{l,l}} {\mathbf{R}_{\bar{\mathbf{w}}_{l}}^{ - 1}} {\bar{\mathbf{D}}_{l,l}} \right) \bm{\beta}_{l}\bm{\beta}_{l}^{T}\nonumber \\
		 & +\! \frac{{{2 T_c}}}{L}\mathrm{tr}\left(  { {\bar{\mathbf{H}}_{l}}^H}{\mathbf{R}_{\bar{\mathbf{w}}_{l}}^{ - 1}}\bar{\mathbf{H}}_{l}{\mathbf{R}_{s,l}}  \right) \mathbf{I}_{2},\label{F_bar_beta_beta}       
	\end{align}}
	\end{subequations}
	with $\bar{\mathbf{B}}_{\theta_{l},l} = {{\dot{\bar{\mathbf{a}}}_{\theta_{l}}}\bar{\mathbf{a}}_l^H + {\bar{\mathbf{a}}_l}\dot{\bar{\mathbf{a}}}_{\theta_{l}}^H}$, $\bar{\mathbf{C}}_{\theta_{l},l} = \left( {\dot{\bar{\mathbf{a}}}_{\theta_{l}} {\mathbf{a}} _l^T + {{\bar {\mathbf{a}} }_l}{\dot{{\mathbf{a}}}_{\theta_{l}}^T}} \right) {{\mathbf{\Psi }}_l}{{\mathbf{G}}_l}$, $\bar{\mathbf{B}}_{\phi_{l},l} = {{\dot{\bar{\mathbf{a}}}_{\phi_{l}}}\bar{\mathbf{a}}_l^H + {\bar{\mathbf{a}}_l}\dot{\bar{\mathbf{a}}}_{\phi_{l}}^H}$, $\bar{\mathbf{C}}_{\phi_{l},l} = \left( {\dot{\bar{\mathbf{a}}}_{\phi_{l}} {\mathbf{a}} _l^T + {{\bar {\mathbf{a}} }_l}{\dot{{\mathbf{a}}}_{\phi_{l}}^T}} \right) {{\mathbf{\Psi }}_l}{{\mathbf{G}}_l}$, $\bar{\mathbf{D}}_{l,l} = {\bar{\mathbf{a}}_l} \bar{\mathbf{a}}_l^H$, and $\bar{\mathbf{H}}_{l} = {\bar{\mathbf{a}}}_{l} {\mathbf{a}} _l^T {{\mathbf{\Psi }}_l}{{\mathbf{G}}_l}$.
\end{prop}
\begin{IEEEproof}
	See Appendix \ref{prop_FIM2_proof}.
\end{IEEEproof}
\begin{rem}
	Note that the first term in \eqref{FIM_def} is non-zero with the deployment of active IRSs. This is attributed to the fact that the noise term $\bar{\mathbf{w}}_{l}$ in \eqref{vec_y_l_irs} contains the information of the unknown parameters ${\bm{\xi}}_{l}$. Conversely, this term is typically null under passive IRSs deployment. In the case of sensing at the BS, the noise term $\mathbf{G}_{l}^{T}\mathbf{\Psi}_{l}\mathbf{E}_{l}\mathbf{\Psi}_{l}{\mathbf{Z}_{l,1}}$ in \eqref{EchoSigAtBS1}, introduced by the active IRS and containing the information of the unknown parameters, is negligible due to the triple reflection over the IRS-target-IRS-BS link. Therefore, the AWGN at the active IRSs contributes to target sensing under the setup with sensing at the IRSs. 
\end{rem}
\subsection{Joint Transmit and Reflective Beamforming Design}
In this subsection, we jointly design the transmit beamforming $\{\mathbf{R}_{s,l}\}$ at the BS and the reflective beamforming $\{\mathbf{\Psi}_{l}\}$ at the IRSs to improve the performance of target estimation. Specifically, we formulate an optimization problem by leveraging the closed-form FIM derived in \eqref{FIM_irs}. Accordingly, we propose an algorithm to solve the formulated problem through alternating optimization.

We aim to minimize the maximum CRB among all IRSs, subject to the constraints on the maximum transmit power at the BS, as well as the maximum transmit power and the maximum power amplification gain constraints at individual IRSs. As such, the optimization problem is formulated as
\begin{subequations}
	\begin{eqnarray}
		(\text{P4}):&{\mathop {\min}\limits_{\{\boldsymbol{\Psi}_{l},\mathbf{R}_{s,l}\} } } \!\!\!\!&\mathop{\max} \limits_{l \in \mathcal{L}}  \quad \overline{\mathrm{CRB}}_{{\bm{\xi}}_{l}}(\mathbf{R}_{s,l},\{\mathbf{\Psi}_{l}\}) \nonumber\\
		&\text{s.t.} \!\!\!\!\!\!\!& \mathrm{tr}\left( \mathbf{\Psi}_{l}\mathbf{G}_{l}\mathbf{R}_{s,l}\mathbf{G}_{l}^{H}\mathbf{\Psi}_{l}^{H} \right) \!+\!  \sigma_{\text{r}}^2 \mathrm{tr}\left(\mathbf{\Psi}_{l}\mathbf{\Psi}_{l}^{H}\right) \nonumber\\
		&&\leq P_{\text{s}}, \forall l\in \mathcal{L}, \label{P4_I_cons1}\\
		&&\eqref{P1_I_cons2},\eqref{P1_I_cons3}. \nonumber
	\end{eqnarray}
\end{subequations}
In problem (P4), \eqref{P4_I_cons1} denotes the transmit power constraints at the active IRSs. Note that problem (P1) is non-convex due to the fact that both the objective function and the transmit power constraint at the IRSs in \eqref{P4_I_cons1} are non-convex. To handle this issue, we alternately optimize the transmit signal covariance $\{\mathbf{R}_{s,l}\}$ at the BS and the reflective beamforming $\{\mathbf{\Psi}_{l}\}$ at the IRSs.
\subsubsection{Optimal Transmit Signal Covariance under Given $\{\mathbf{\Psi}_{l}\}$}
Under given $\{\mathbf{\Psi}_{l}\}$, the optimization of transmit beamforming $\{\mathbf{R}_{s.l}\}$ is reformulated as
\begin{align}
	\begin{array}{*{20}{lcl}}
		(\text{P5}): & {\mathop {\min}\limits_{\{\mathbf{R}_{s,l}\} } } & \mathop{\max} \limits_{l \in \mathcal{L}}  \quad \mathrm{tr}(\bar{\mathbf{F}}_{l}^{-1}) \nonumber \\
		             & \text{s.t.}                                      & \eqref{P1_I_cons1}-\eqref{P1_I_cons3}.\nonumber
	\end{array}
\end{align}
To solve the non-convex problem (P5), we first introduce an auxiliary $\bar{\kappa}$ to transform it into the following equivalent problem:
\begin{subequations}
	\begin{eqnarray}
		(\text{P5.1}):&{\mathop {\min}\limits_{ \{\mathbf{R}_{s,l}\},\bar{\kappa}  } } &  \bar{\kappa} \nonumber\\
		&\text{s.t.} & \mathrm{tr}(\bar{\mathbf{F}}_{l}^{-1}) \leq \bar{\kappa}, \forall l\in \mathcal{L} \label{P5_1_cons4},\\
		&& \eqref{P1_I_cons1}-\eqref{P1_I_cons3}.\nonumber
	\end{eqnarray}
\end{subequations}
From the expression of FIM in \eqref{FIM_irs}, we note that $\bar{\mathbf{F}}_{l}$ is a linear function of $\{\mathbf{R}_{s,l}\}$ and thus $\mathrm{tr}(\bar{\mathbf{F}}_{l}^{-1})$ is a convex function of $\{\mathbf{R}_{s,l}\}$. Therefore, the constraints in \eqref{P5_1_cons4} are convex. As a result, problem (P5.1) is a convex SDP, which can be optimally solved by CVX.
\subsubsection{Reflective Beamforming Optimization under Given $\{\mathbf{R}_{s,l}\}$}
Under given transmit beamforming $\{\mathbf{R}_{s,l}\}$, the optimization of reflective beamforming $\{\boldsymbol{\Psi}_{l}\}$ is reformulated as
\begin{align}
	 & \begin{array}{*{20}{lcl}}
		(\text{P6}): & {\mathop {\min}\limits_{\{\boldsymbol{\Psi}_{l}\} } } & \mathop{\max} \limits_{l \in \mathcal{L}}  \quad \mathrm{tr}(\bar{\mathbf{F}}_{l}^{-1}) \nonumber \\
		             & \text{s.t.}                                           & \eqref{P4_I_cons1}.\nonumber
	\end{array}
\end{align}
Note that the FIM $\bar{\mathbf{F}}_{l}$ depends only on the reflective beamforming $\boldsymbol{\Psi}_{l}$ at IRS $l$. Similar to problem (P3), problem (P6) can also be equivalently decomposed into $L$ subproblems as
\begin{align}
		\begin{array}{*{20}{lcl}}
		(\text{P6.$l$.1}):&{\mathop {\min}\limits_{\boldsymbol{\Psi}_{l} } } & \mathrm{tr}(\bar{\mathbf{F}}_{l}^{-1}) \nonumber\\
		&\text{s.t.} & \eqref{P4_I_cons1}.\nonumber
		\end{array}
\end{align}
This problem is highly non-convex due to the fact that the transmit power constraints at IRSs in \eqref{P1_I_cons1} and the elements in FIM are non-convex functions of $\{\boldsymbol{\Psi}_{l}\}$. By introducing anxiliary variables $\{\kappa_i\}_{i=1}^{4}$, problem (\text{P6.$l$.1}) is equivalent to
\begin{subequations}
	\begin{eqnarray}
		\!\!\!\!(\text{P6.$l$.2}):&\!\!\!\!\!{\mathop {\min}\limits_{\boldsymbol{\Psi}_{l},\{\kappa_i\}_{i=1}^{4}  } } & \sum\nolimits_{i=1}^{4} \kappa_i \nonumber\\
		&\text{s.t.} &\!\!\!\!\! \left[\begin{array}{cc}
				\bar{\mathbf{F}}_{l} & \mathbf{e}_{i} \\
				\mathbf{e}_{i}^{T}   & \kappa_i
			\end{array}\right] \succeq 0, i \in \{1, \ldots, 4\}, \label{Schur_Com2}\\
		&&\eqref{P4_I_cons1},\nonumber
	\end{eqnarray}
\end{subequations}
where constraint \eqref{Schur_Com2} is derived by utilizing the Schur complement. Note that problem (P6.$l$.2) is still non-convex due to the non-convex constraints in \eqref{P4_I_cons1} and \eqref{Schur_Com2}, and then we adopt SDR and SCA techniques to transform these constraints into a convex form. First, we approximate FIM in \eqref{FIM_irs} based the following lemma.

\begin{lem}\label{bar_FIM_appr_lem}
	With a given local point $\mathbf{\Theta}_{l}^{(i)}$, the FIM in \eqref{FIM_irs} is approximated as
	\begin{align}
		\hat{\bar{\mathbf{F}}}_{l} = \left[\begin{array}{ccc}
				\hat{\bar{{F}}}_{\theta_{l},\theta_l}                  & \hat{\bar{{F}}}_{\theta_{l},\phi_{l}}                & \hat{\bar{\mathbf{F}}}_{\theta_{l},\bm{\beta}_{l}}       \\
				\hat{\bar{{F}}}_{\theta_{l},\phi_{l}}^{T}              & \hat{\bar{{F}}}_{\phi_{l},\phi_{l}}                  & \hat{\bar{\mathbf{F}}}_{\phi_{l},\bm{\beta}_{l}}         \\
				\hat{\bar{\mathbf{F}}}_{\theta_{l},\bm{\beta}_{l}}^{T} & \hat{\bar{\mathbf{F}}}_{\phi_{l},\bm{\beta}_{l}}^{T} & \hat{\bar{\mathbf{F}}}_{{\bm{\beta}_{l}}.\bm{\beta}_{l}}
			\end{array}\right],\label{FIM_bs_trans}
	\end{align}
	in which
		\begin{subequations}
		{\small\begin{align}
			&\hat{\bar{F}}_{\varrho,\varpi}                       =  {\frac{4T_{c}}{L}}\sigma _r^4{\left| {{\beta _l}} \right|^4}  {\mathrm{tr}}\left( {\mathbf{R}_{\bar{\mathbf{w}}_{l}}^{ - 1}}\bar{\mathbf{B}}_{\varrho,l}{\mathbf{R}_{\bar{\mathbf{w}}_{l}}^{ - 1}}\bar{\mathbf{B}}_{\varpi,l} \right) \left( 2 \mathrm{tr}\left(\mathbf{\Theta}_{l}^{(i)}\right)\mathrm{tr}\left(\mathbf{\Theta}_{l}\right)\right.  \nonumber \\
			& \left. - \mathrm{tr}^{2}\left(\mathbf{\Theta}_{l}^{(i)}\right)  \right) + \frac{{{2 T_c}}}{L}{\left| {{\beta _l}} \right|^2} \mathrm{tr}\left(\bar{\nabla}_{\varrho,\varpi}^{T}(\mathbf{\Theta}_{l}^{(i)}) \left(\mathbf{\Theta}_{l}-\mathbf{\Theta}_{l}^{(i)}\right)  \right) \nonumber\\
			& + \frac{{{2 T_c}}}{L}{\left| {{\beta _l}} \right|^2} \bar{Q}_{\varrho,\varpi}(\mathbf{\Theta}_{l}^{(i)}),\label{hat_F_theta_theta}  \\
			&\hat{\bar{\mathbf{F}}}_{\varrho,\bm{\beta}_{l}}        = \frac{{4{T_c}}}{L}\sigma _r^4{\left| {{\beta _l}} \right|^2}  {\mathrm{tr}}\left({\mathbf{R}_{\bar{\mathbf{w}}_{l}}^{ - 1}}\bar{\mathbf{B}}_{\varrho,l}{\mathbf{R}_{\bar{\mathbf{w}}_{l}}^{ - 1}}  {\bar{\mathbf{D}}_{l,l}} \right)\left( 2 \mathrm{tr}\left(\mathbf{\Theta}_{l}^{(i)}\right)\mathrm{tr}\left(\mathbf{\Theta}_{l}\right) \right. \nonumber \\
			&\left. - \mathrm{tr}^{2}\left(\mathbf{\Theta}_{l}^{(i)}\right)  \right) [\Re \left( {{\beta _l}} \right),\Im \left( {{\beta _l}} \right)] + \frac{{{2 T_c}}}{L}\Re \left( \beta_{l} \left(  \mathrm{tr} \left(\bar{\nabla}_{\varrho,\bm{\beta}_{l}}^{T}(\mathbf{\Theta}_{l}^{(i)}) \right.\right.\right.\nonumber\\
			& \left.\left.\left. \left(\mathbf{\Theta}_{l} - \mathbf{\Theta}_{l}^{(i)}\right)\right) + \bar{Q}_{\varrho,\bm{\beta}_{l}}(\mathbf{\Theta}_{l}^{(i)}) \right) [1, j]\right),\label{hat_F_theta_beta}  \\
			&\hat{\bar{\mathbf{F}}}_{{\bm{\beta}_{l}},\bm{\beta}_{l}}  = \frac{{{4T_c}}}{L} \sigma_r^4  {\mathrm{tr}}\left( {\mathbf{R}_{\bar{\mathbf{w}}_{l}}^{ - 1}} {\bar{\mathbf{D}}_{l,l}} {\mathbf{R}_{\bar{\mathbf{w}}_{l}}^{ - 1}} {\bar{\mathbf{D}}_{l,l}} \right)\left( 2 \mathrm{tr}\left(\mathbf{\Theta}_{l}^{(i)}\right)\mathrm{tr}\left(\mathbf{\Theta}_{l}\right) \right.\nonumber  \\
			& \left. - \mathrm{tr}^{2}\left(\mathbf{\Theta}_{l}^{(i)}\right)  \right) \bm{\beta}_{l}\bm{\beta}_{l}^{T} + \frac{{{2 T_c}}}{L} \left( \mathrm{tr}\left(\bar{\nabla}_{{\bm{\beta}_{l}},\bm{\beta}_{l}}^{T}(\mathbf{\Theta}_{l}^{(i)})\left(\mathbf{\Theta}_{l} - \mathbf{\Theta}_{l}^{{i}}\right)  \right) \right.\nonumber\\
			&\left.+ \bar{Q}_{{\bm{\beta}_{l}},\bm{\beta}_{l}}(\mathbf{\Theta}_{l}^{(i)}) \right) \mathbf{I}_{2}.\label{hat_F_beta_beta}
		\end{align}}
	\end{subequations}
\end{lem}
\begin{IEEEproof}
	See Appendix \ref{bar_FIM_appr_proof}.
\end{IEEEproof}
Then, constraint \eqref{P4_I_cons1} is equivalent to
\begin{align}
	\mathrm{tr}\left(\mathbf{G}_{l}\mathbf{R}_{s,l}\mathbf{G}_{l}^{H}\mathrm{Diag}\left(\mathbf{\Theta}_{l}\right)\right) +  \sigma_{\text{r}}^2 \mathrm{tr}\left(\mathbf{\Theta}_{l}\right) \leq P_{\text{s}}, \forall l\in \mathcal{L}, \label{P4_I_cons1_trans}
\end{align}
which is convex with respect to $\mathbf{\Theta}_{l}$. As a result, problem (P6.$l$.2) is transformed into the following problem at iteration $i$ of SCA.
\begin{subequations}
	\begin{eqnarray}
		\!\!\!(\text{P6.$l$.3.$i$}):&\!\!\!\!\!\! {\mathop {\min}\limits_{\mathbf{\Theta}_{l},\{\kappa_i\}_{i=1}^{4}  } } \!\!\!\!& \sum\nolimits_{i=1}^{4} \kappa_i \nonumber\\
		&\text{s.t.} &\!\!\!\!\!\! \left[\begin{array}{cc}
				\hat{\bar{\mathbf{F}}}_{l} & \mathbf{e}_{i} \\
				\mathbf{e}_{i}^{T}         & \kappa_i
			\end{array}\right] \succeq 0, i \in \{1, \ldots, 4\}, \label{Schur_Com2}\\
		&&\mathrm{rank}(\mathbf{\Theta}_{l} ) = 1, \\
		&&\eqref{P4_I_cons1_trans}.\nonumber
	\end{eqnarray}
\end{subequations}
By dropping the rank-one constraint in problem (P6.$l$.3.$i$), we obtain the relaxed version of (P6.$l$.3.$i$) as (SDR6.$l$.3.$i$). Note that problem (SDR6.$l$.3.$i$) is a convex problem which can be efficiently solved by CVX. Let $\mathbf{\Theta}_{l}^{*(i)}$ denote the obtained solution to problem (P6.$l$.3.$i$) at the $i$-th iteration, which is then updated as $\mathbf{\Theta}_{l}^{*(i+1)}$. Note that the objective value achieved by $\mathbf{\Theta}_{l}^{*(i+1)}$ is always no greater than that by $\mathbf{\Theta}_{l}^{*(i)}$. It means that the achieved CRB is monotonically non-increasing after each iteration of SCA. Besides, the optimal value of problem (P6.$l$.2) is lower bounded due to the non-negativity of the entries of FIM. Thus, the convergence of the SCA is guaranteed, and we denote $\mathbf{\Theta}_{l}^{*}$ as the corresponding converged solution. Note that $\mathbf{\Theta}_{l}^{*}$ is generally not rank-one. By adopting the Gaussian randomization, we obtain an efficient rank-one solution to (P6.$l$.2) based on the obtained $\mathbf{\Theta}_{l}^{*}$. More specifically, we generate a number of random vectors ${\mathbf{r}}_{l} \sim \mathcal{CN}\left(\mathbf{0},\mathbf{I}_N\right)$, and then construct a number of rank-one solutions as
\begin{align}\label{GR_construct}
	\boldsymbol{\psi}_{l} = \left(\boldsymbol{\Theta}_{l}^{*}\right)^{\frac{1}{2}}\mathbf{r}_{l}.
\end{align}
With these generated $\boldsymbol{\psi}_{l}$'s, we verify whether the maximum amplitude of the elements in $\boldsymbol{\psi}_{l}$ exceeds the maximum amplification gain $a_{\text{max}}$. If so, we normalize $\boldsymbol{\psi}_{l} = a_{\text{max}}\frac{\boldsymbol{\psi}_{l}}{\max(|\boldsymbol{\psi}_{l}|)}$, where $\max(|\boldsymbol{\psi}_{l}|)$ represents the maximum amplitude value of the elements in $\boldsymbol{\psi}_{l}$. Then, we find the desirable solution of $\boldsymbol{\psi}_{l}$ that minimizes $\overline{\mathrm{CRB}}_{{\bm{\xi}}_{l}}(\mathbf{R}_{s,l},\mathbf{\Psi}_{l})$ meanwhile satisfying the constraint \eqref{P4_I_cons1} among all randomly generated $\boldsymbol{\psi}_{l}$'s.  As a result, problem (P6) is finally solved.

In summary, the alternating optimization-based algorithm for solving (P4) is implemented by solving problems (P5) and (P6) alternately. Similar to problem (P1), the proposed alternative optimization-based algorithm leads to monotonically non-increasing max-CRB values for problem (P4) throughout iterations, thus securing convergence. Additionally, it is suitable to be implemented in a distributed manner.

\section{Numerical Results}
\begin{figure}[tbp]
	\setlength{\abovecaptionskip}{-0pt}
	\setlength{\belowcaptionskip}{-12pt}
	\centering
	\includegraphics[width= 0.4\textwidth]{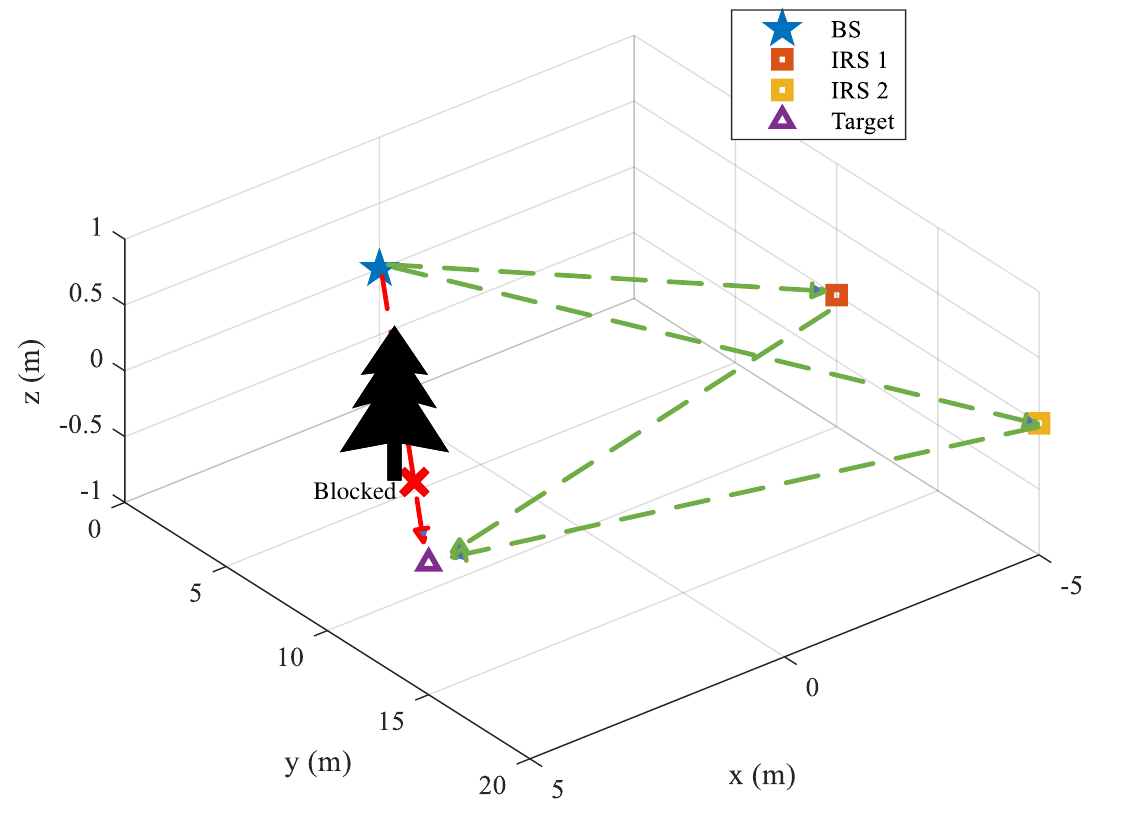}
	\DeclareGraphicsExtensions.
	\caption{\color{black}The location topology.}
	\label{Topology}
\end{figure}

This section provides numerical results to validate the effectiveness of our proposed design. In the simulation, we adopt the Rician fading channel model with the K-factor being $5$ dB for channels between the BS and IRSs. The channels between IRSs and the target are assumed to be LoS channels. Additionally, we set the noise power as $\sigma_{\text{r}}^{2}= \sigma_{\text{b}}^{2} = -80$ dBm, and the radar dwell time as $T_c = 100$ time symbols.
In particular, we consider a scenario with one BS, two active IRSs, and one target as shown in Fig. \ref{Topology}. The BS and two IRSs are located at $(0,0,0)$ meters~(m),  $(-5,10,0)$ m, and $(-5,20,0)$ m, respectively. The target is located at $(5,15,0)$ m. To better illustrate the superiority of our proposal, we adopt the following two benchmarks for comparison.

\textbf{Transmit beamforming (BF) only}: The IRSs implement random reflection coefficients as $\bm{\psi}_{l} = a_{\text{max}}e^{j\mathrm{arg}\left\{\mathbf{r}\right\}}, \forall l\in \mathcal{L}$, where ${\mathbf{r}} \sim \mathcal{CN}\left(\mathbf{0},\mathbf{I}_N\right)$. Accordingly, we only optimize the transmit beamforming at the BS by solving problems (P2) and (P5) in Sections II-C-1) and III-C-1), respectively.

\textbf{Reflective BF only}: The BS adopts the isotropic transmission by setting $\mathbf{R}_{s,l}=\frac{P_{\text{t}}}{M}\mathbf{I}_{M}, \forall l\in \mathcal{L}$. Then, we optimize the reflective beamforming at all IRSs by solving problems (P3) and (P6) in Sections II-C-2) and III-C-2).

\textbf{Passive IRS-enabled design}: This benchmark refers to passive IRSs deployment. In this scheme, we optimize the reflective beamforming at all IRSs with the constraint $[\mathbf{\Theta}_{l}]_{n,n} = 1$, and the optimization scheme is similar to that in \cite{fang2023multiirsenabled}.
\subsection{Sensing at BS}
\begin{figure}[tbp]
	\setlength{\abovecaptionskip}{-0pt}
	\setlength{\belowcaptionskip}{-12pt}
	\centering
	\includegraphics[width= 0.35\textwidth]{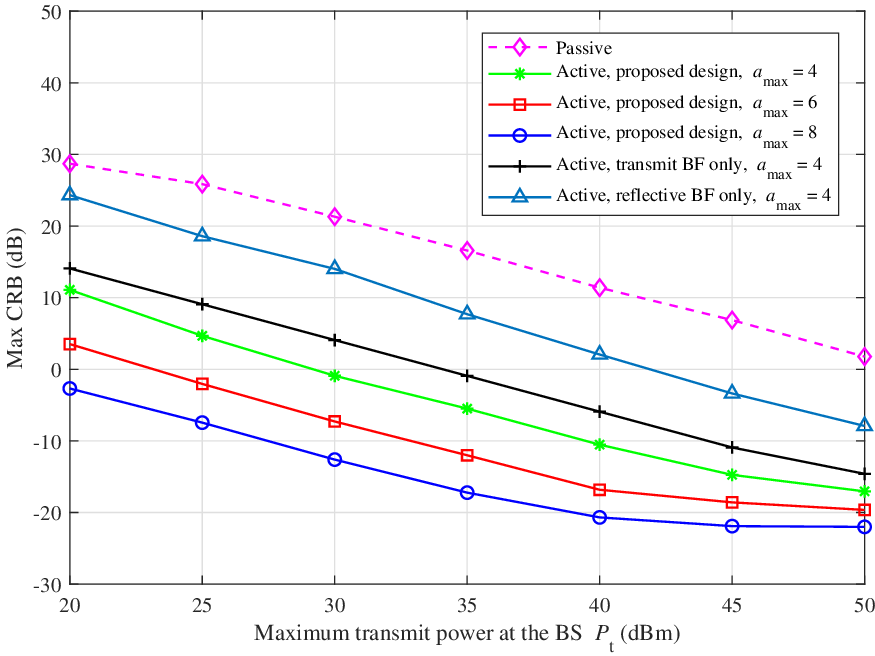}
	\DeclareGraphicsExtensions.
	\caption{\color{black}The achieved max-CRB versus the maximum transmit power $P_{\text{t}}$ at the BS with $P_{\text{s}} = 0.1$ W, $M=16$, and $N_v = N_h = 4$.}
	\label{CRBVsPt}
\end{figure}
First, we consider the case of sensing at the BS. Fig.~\ref{CRBVsPt} plots the achieved max-CRB versus the maximum transmit power $P_{\text{t}}$ at the BS. First, it is observed that our proposed design outperforms other benchmark schemes, and the max-CRB performance achieved by active IRSs outperforms that of passive IRSs by a significant margin. This clearly shows the benefit of deploying active IRSs for wireless sensing. Besides, the max-CRB achieved by the `transmit BF only' benchmark is lower than that of the `reflective BF only' one. This indicates that transmit beamforming plays a more prominent role in the considered sensing system. In particular, a delicate design of transmit beamforming to direct beams toward multiple active IRSs is a rule of thumb for establishing high-quality sensing links. Furthermore, it is also observed that the higher the maximum amplification gain at the IRSs, the better sensing performance can be achieved. This is because a looser maximum amplification gain constraint of the elements at IRSs implies more degrees of freedom in reflective beamforming.

\begin{figure}[t]
	\setlength{\abovecaptionskip}{-0pt}
	\setlength{\belowcaptionskip}{-12pt}
	\centering
	\includegraphics[width= 0.35\textwidth]{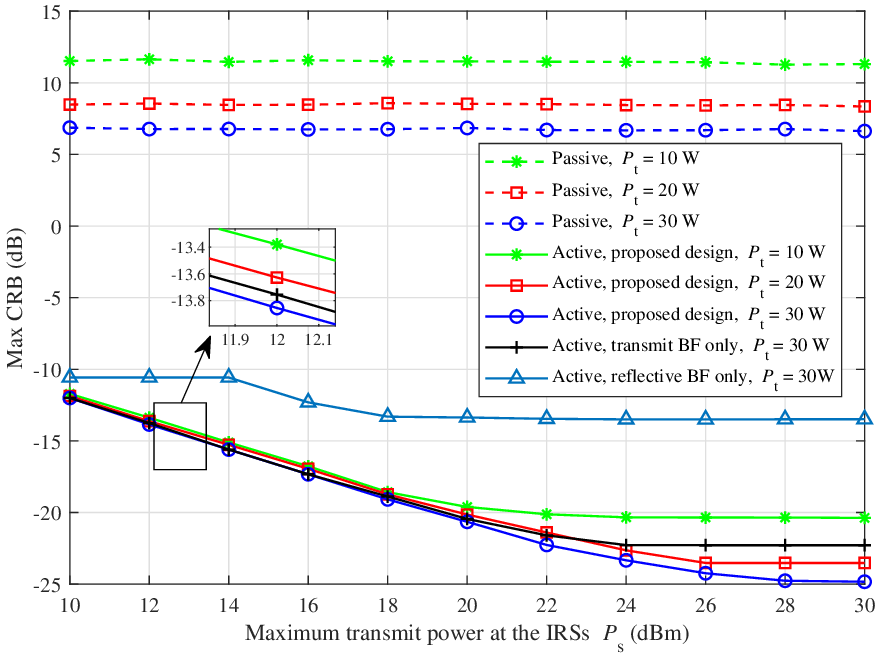}
	\DeclareGraphicsExtensions.
	\caption{\color{black}The achieved max-CRB versus the maximum transmit power $P_{\text{s}}$ at the IRSs with $M=16$, $N_v = N_h = 4$, and $a_{\text{max}}=8$.}
	\label{CRBVsPs}
\end{figure}
Fig.~\ref{CRBVsPs} shows the achieved max-CRB versus the maximum transmit power $P_{\text{s}}$ at the IRSs. In the low $P_{\text{s}}$ regime, it is observed that the CRBs under different $P_{\text{t}}$ are almost the same and decrease as $P_{\text{s}}$ increases. This phenomenon occurs because the received echo signal power is primarily constrained by the maximum transmit power budget at the IRSs. Subsequently, in the high $P_{\text{s}}$ regime, it is observed that the CRBs remain constant. This is attributed to the fact that the maximum transmit power budget at the IRS is large enough, and the received echo signal power is mainly constrained by the transmit power at the BS instead.

\subsection{Sensing at IRSs}
\begin{figure}[t]
	\setlength{\abovecaptionskip}{-0pt}
	\setlength{\belowcaptionskip}{-12pt}
	\centering
	\includegraphics[width= 0.35\textwidth]{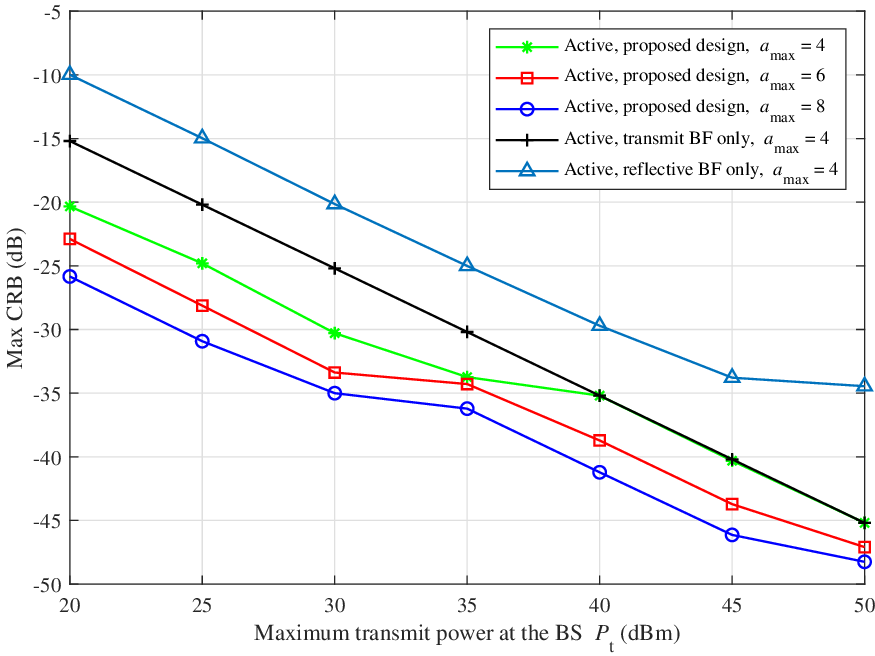}
	\DeclareGraphicsExtensions.
	\caption{\color{black}The achieved max-CRB versus the maximum transmit power $P_{\text{t}}$ at the BS with $P_{\text{s}} = 0.1$ W, $M=16$, and $N_v = N_h = 4$.}
	\label{CRBVsPtIRS}
\end{figure}
Next, we consider the sensing at the IRS cases in which we set the number of sensors at the IRSs as $\bar{N}_v = \bar{N}_h = 4$. Fig.~\ref{CRBVsPtIRS} illustrates the achieved max-CRB versus the maximum transmit power $P_{\text{t}}$ at the BS. First, it is observed that our proposed design outperforms other benchmark schemes. Similar to the case of sensing at the BS, the max-CRB achieved by the `transmit BF only' benchmark is lower than that of the `reflective BF only' one. Besides, it is also shown that the `transmit BF only' performs the same as the proposed design in the high $P_{\text{t}}$ regime. This indicates that transmit beamforming plays a more prominent role when $P_{\text{t}}$ is high. Furthermore, by comparing Fig. \ref{CRBVsPt} and Fig. \ref{CRBVsPtIRS}, it is observed that the sensing at the IRSs setup outperforms the sensing at the BS one, which shows the advantages of deploying dedicated sensors at the IRSs. This is also because the sensing signal only suffers two hops of reflections compared to three hops for sensing at the BS. However, employing sensing at the IRSs needs to equip additional sensors at them thus increasing the cost, which demonstrates a tredeoff between performance and cost in practical IRSs deployment.

\begin{figure}[t]
	\setlength{\abovecaptionskip}{-0pt}
	\setlength{\belowcaptionskip}{-8pt}
	\centering
	\includegraphics[width= 0.35\textwidth]{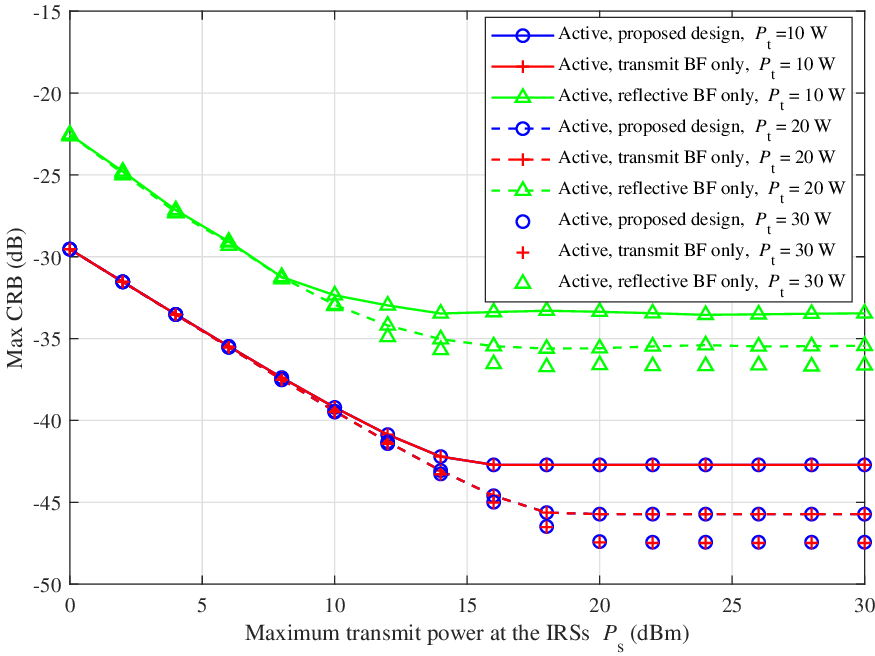}
	\DeclareGraphicsExtensions.
	\caption{\color{black}The achieved max-CRB versus the maximum transmit power $P_{\text{s}}$ at the IRSs with $M=16$, $N_v = N_h = 4$, and $a_{\text{max}}=8$.}
	\label{CRBVsPsIRS}
\end{figure}
Fig.~\ref{CRBVsPsIRS} shows the achieved max-CRB versus the maximum transmit power $P_{\text{s}}$ at the IRSs. In the low $P_{\text{s}}$ regime, it is observed that the CRBs under different $P_{\text{t}}$ are almost the same and decrease as $P_{\text{s}}$ increases. This phenomenon occurs because the received echo signal power is primarily constrained by the maximum transmit power budget at the IRSs. It is also observed that the `transmit BF only' performs the same as proposed design, which coincides with Fig. \ref{CRBVsPtIRS}. This provides further verification of the critical role of transmit beamforming at the BS.
\subsection{Sensing at BS versus Sensing at IRSs}
\begin{figure}[t]
	\setlength{\abovecaptionskip}{-0pt}
	\setlength{\belowcaptionskip}{-17pt}
	\centering
	\includegraphics[width= 0.35\textwidth]{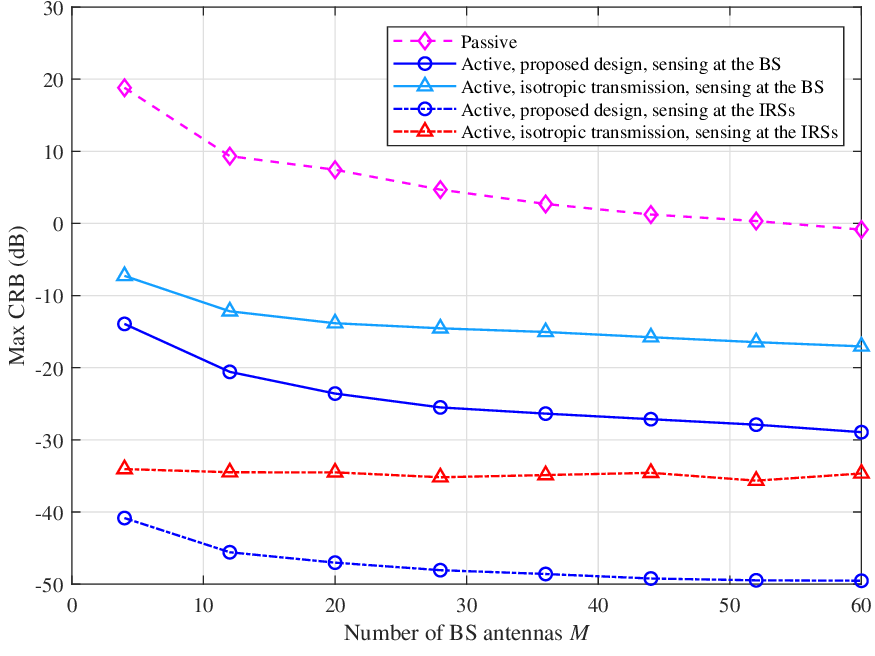}
	\DeclareGraphicsExtensions.
	\caption{\color{black}The achieved max-CRB versus the number of antennas $M$ at the BS with $P_{\text{t}} = 20$ W, $P_{\text{s}} = 0.1$ W, $N_v = N_h = 4$, and $a_{\text{max}}=8$.}
	\label{CRBVsBSantenna}
\end{figure}
Finally, we compare the cases involving sensing at the BS and at the IRSs. Fig.~\ref{CRBVsBSantenna} plots the achieved max-CRB versus the number of antennas $M$ at the BS. When sensing at the BS, it is observed that the CRB decreases as  $M$ increases, but the performance saturates as $M$ further increases. In particular, when the number of antennas, $M$, starts to increase from a small number, the received echo signal power is enhanced rapidly due to the enlarged array gain. However, when $M$ continues to increase, the array gain is no longer the limiting factor of sensing performance. Instead, the maximum transmit power budgets at the BS and IRSs limit the received echo signal power. This leads to a diminishing performance gain in terms of max-CRB when more antennas are deployed at the BS. Regarding sensing at the IRSs, the CRB achieved by our proposed design has a similar trend to sensing at the BS, albeit with significantly better performance. However, it is shown that the CRB performance achieved by the isotropic transmission method is independent of $M$. This is because the received signal power at the IRSs from the BS-IRS link is independent of the number of antennas at the BS under isotropic transmission and the received echo signal power by the sensors at the IRSs from the IRS-target-IRS link is also unaffected by the number of antennas at the BS.
\vspace{-0.1cm}
\section{Conclusion}
This paper investigated multi-active-IRS-assisted cooperative sensing, where multiple active IRSs are deployed to provide multi-view sensing. In particular, we considered two different configurations of active IRSs, one with dedicated sensors for signal reception and processing and the other without. In the first configuration, the BS was used to receive the echo signal and estimate the target's DoA with respect to each IRS. In the second configuration, each IRS uses its dedicated sensors to receive and estimate the target's DoA with respect to itself. We proposed a multi-active-IRS time division sensing framework and then derived the closed-form CRBs for estimation of target parameters. Then, we proposed an efficient joint transmit and reflective beamforming design to minimize the maximum CRB among all IRSs subject to practical constraints. Numerical results demonstrated the effectiveness of our proposed design, and showed that active IRSs outperform passive ones by a significant margin for target sensing. Furthermore, it was shown that a higher maximum transmit power budget and maximum amplification gain constraint at the IRSs can lead to improved sensing performance, especially when the transmit power budget at the BS becomes large. Additionally, it indicated that the design of transmit beamforming is more critical than that of reflective beamforming.

\vspace{-2mm}
\begin{appendices}
	\section{Proof of Proposition \ref{prop_FIM_bs}}\label{prop_FIM1_proof}
	We define $\mathbf{D}_{N} = \mathrm{diag}\{\mathbf{d}_{N}\}$ with $\mathbf{d}_{N} = [0,1,\cdots,N-1]$. Then, we derive the derivative of ${\mathbf{a}}_{l}$ with respect to $\theta_{l}$ as
	\begin{align}
		 & \dot{{\mathbf{a}}}_{\theta_{l}} = \frac{{\partial {{\mathbf{a}}_l}}}{{\partial {\theta _l}}} = \frac{{\partial {{{\mathbf{a}}}_v}({\theta _l}) \otimes {{{\mathbf{a}}}_h}({\theta _l},{\phi _l})}}{{\partial {\theta _l}}}\nonumber                                                                 \\
		 & = \frac{{\partial {{{\mathbf{a}}}_v}({\theta _l})}}{{\partial {\theta _l}}} \otimes {{\mathbf{a}}_h}({\theta _l},{\phi _l}) + {{\mathbf{a}}_v}({\theta _l}) \otimes \frac{{\partial {{{\mathbf{a}}}_h}({\theta _l},{\phi _l})}}{{ \partial {\theta _l}}}\nonumber \\
		 & =- j\frac{{2\pi {{d}_v}}}{\lambda }\sin ({\theta _l})\left( {{\mathbf{d}_{{{ N}_v}}} \otimes {\mathbf{1}_{{{N}_h}}}} \right) \circ {{\mathbf{a}}_l} \nonumber                                                                                                     \\
		 & +j\frac{{2\pi {{ d}_h}}}{\lambda }\cos ({\theta _l})\cos ({\phi _l})\left( {{\mathbf{1}_{{{ N}_v}}} \otimes {\mathbf{d}_{{{ N}_h}}}} \right) \circ {{\mathbf{a}}_l} \nonumber                                                                                     \\
		 & = \bm{\zeta}_{\theta_l} \circ {{\mathbf{a}}_l} = \bm{Z}_{\theta_l} {{\mathbf{a}}_l},\label{De_a_theta}
	\end{align}
	where $\bm{Z}_{\theta_l} = \mathrm{diag}(\bm{\zeta}_{\theta_l})$ with $\bm{\zeta}_{\theta_l} = j\frac{{2\pi {{ d}_h}}}{\lambda }\cos ({\theta _l})\cos ({\phi _l})\left( {{\mathbf{1}_{{{ N}_v}}} \otimes {\mathbf{d}_{{{ N}_h}}}} \right) - j\frac{{2\pi {{ d}_v}}}{\lambda }\sin ({\theta _l})\left( {{\mathbf{d}_{{{ N}_v}}} \otimes {\mathbf{1}_{{{ N}_h}}}} \right) $ and $\mathbf{1}_{N} \in \mathbb{R}^{N\times1}$ with all elements being one. Similarly, the derivative of ${\mathbf{a}}_{l}$ with respect to $\phi_{l}$ is derived by
		\begin{align}
			&\dot{{\mathbf{a}}}_{\phi_{l}} = \frac{{\partial {{\mathbf{a}}_l}}}{{\partial {\phi_l}}} = \frac{{\partial {{{\mathbf{a}}}_v}({\theta _l}) \otimes {{{\mathbf{a}}}_h}({\theta _l},{\phi _l})}}{{\partial {\phi _l}}}\nonumber\\
			&= \frac{{\partial {{{\mathbf{a}}}_v}({\theta _l})}}{{\partial {\phi _l}}} \otimes {{\mathbf{a}}_h}({\theta _l},{\phi _l}) + {{\mathbf{a}}_v}({\theta _l}) \otimes \frac{{\partial {{{\mathbf{a}}}_h}({\theta _l},{\phi _l})}}{{ \partial {\phi _l}}}\nonumber\\
			&= \! - j\frac{{2\pi {{ d}_h}}}{\lambda }\sin ({\theta _l})\! \sin ({\phi _l})\! \left( {{\mathbf{1}_{{{ N}_v}}}\!\! \otimes {\mathbf{d}_{{{ N}_h}}}} \right)\! \circ \! \left( {{{{\mathbf{a}}}_v}({\theta _l})\!\! \otimes {{{\mathbf{a}}}_h}({\theta _l},{\phi _l})} \right)\nonumber\\
			&= \bm{\zeta}_{\phi_l} \circ {{\mathbf{a}}_l} = \bm{Z}_{\phi_l} {{\mathbf{a}}_l},\label{De_a_phi}
		\end{align}
		where $\bm{Z}_{\phi_l} = \mathrm{diag}(\bm{\zeta}_{\phi_l})$ with $\bm{\zeta}_{\phi_l} =- j\frac{{2\pi {{ d}_h}}}{\lambda }\sin ({\theta _l})\sin ({\phi _l})\left( {{\mathbf{1}_{{{ N}_v}}} \otimes {\mathbf{d}_{{{ N}_h}}}} \right)$. Next, we derive the entry ${{F}}_{\theta_{l},\theta_l}$ in FIM. According to the definition of FIM, ${{F}}_{\theta_{l},\theta_l}$ is given by 
		{\small	\begin{align}
				\!\!{{F}}_{\theta_{l},\theta_l} \!=\! \mathrm{tr}\left(\mathbf{R}_{{\mathbf{y}}_{l}}^{-1}\frac{\partial \mathbf{R}_{{\mathbf{y}}_{l}}}{\partial \theta_{l}}\mathbf{R}_{{\mathbf{y}}_{l}}^{-1}\frac{\partial \mathbf{R}_{{\mathbf{y}}_{l}}}{\partial \theta_{l}}\right) \!\!+\! 2\Re\left(\frac{\partial {\bm{\eta}}_{l}^{H}}{\partial \theta_{l} } \mathbf{R}_{{\mathbf{y}}_{l}}^{-1} \frac{\partial {\bm{\eta}}_{l}}{\partial \theta_{l} }\right). \label{FIM_bs_theta_theta}
			\end{align}}
By substituting \eqref{R_y_bar_bs} into \eqref{FIM_bs_theta_theta}, we have
			\begin{align}
				&{\mathrm{tr}}\left( {{\mathbf{R}}_{{{ {\mathbf{y}} }_l}}^{ - 1}\frac{{\partial {{\mathbf{R}}_{{{ {\mathbf{y}} }_l}}}}}{{\partial {\theta _l}}}{\mathbf{R}}_{{{ {\mathbf{y}} }_l}}^{ - 1}\frac{{\partial {{\mathbf{R}}_{{{ {\mathbf{y}} }_l}}}}}{{\partial {\theta _l}}}} \right) = 0. \label{F_bs_bar_theta_theta_1_1}
			\end{align}
	Then, we substitute \eqref{vec_sig_bs} into \eqref{FIM_bs_theta_theta} and obtain
	{\small
	\begin{align}
		 & \Re\left(\frac{\partial {\bm{\eta}}_{l}^{H}}{\partial \theta_{l} } \mathbf{R}_{{\mathbf{y}}_{l}}^{-1} \frac{\partial {\bm{\eta}}_{l}}{\partial \theta_{l} }\right) =  \nonumber                                                                                                                                                                                                                                                                                                     \\
		 & \Re \left( {\sum\limits_{t = \frac{(l - 1)T_c}{L} + 1}^{l{\frac{T_c}{L}}}\!\!\!\!\!\!\! {{{\left( \!\!{\frac{{\partial \mathbf{G}_{l}^{T}\mathbf{\Psi}_{l}{\mathbf{E}}_{l}\mathbf{\Psi}_{l}\mathbf{G}_{l} {{\mathbf{s}}_l}[t]}}{{\partial {\theta _l}}}} \right)}^H}\!\!\!\!{\mathbf{R}_{{\mathbf{w}}_{l}}^{ - 1}}\frac{{\partial \mathbf{G}_{l}^{T}\mathbf{\Psi}_{l}{\mathbf{E}}_{l}\mathbf{\Psi}_{l}\mathbf{G}_{l}{{\mathbf{s}}_l}[t]}}{{\partial {\theta _l}}}} } \right)  \nonumber \\
		 & = {\left| {{\beta _l}} \right|^2}\Re \left( \sum\limits_{t = \frac{(l - 1)T_c}{L} + 1}^{l{\frac{T_c}{L}}}  {\mathbf{s}}_l^H[t]{\mathbf{G}}_l^H{\mathbf{\Psi }}_l^H{{\left( {\dot{{\mathbf{a}}}_{\theta_{l}} {\mathbf{a}} _l^T + {{ {\mathbf{a}} }_l}{\dot{{\mathbf{a}}}_{\theta_{l}}^T}} \right)}^H} \times \right.\nonumber                                                                                                                                                                \\
		 & \left. \mathbf{\Psi}_{l} \mathbf{G}_{l}^{*} {\mathbf{R}_{{\mathbf{w}}_{l}}^{ - 1}} \mathbf{G}_{l}^{T}\mathbf{\Psi}_{l} \left( {\dot{{\mathbf{a}}}_{\theta_{l}} {\mathbf{a}} _l^T + {{ {\mathbf{a}} }_l}{\dot{{\mathbf{a}}}_{\theta_{l}}^T}} \right){{\mathbf{\Psi }}_l}{{\mathbf{G}}_l}{{\mathbf{s}}_l}[t]  \right)\nonumber                                                                                                                                                        \\
		 & ={\left| {{\beta _l}} \right|^2}\frac{{{T_c}}}{L}\mathrm{tr}\left(  { \mathbf{C}_{\theta_{l},l}^H} {\mathbf{R}_{{\mathbf{w}}_{l}}^{ - 1}} \mathbf{C}_{\theta_{l},l} {\mathbf{R}_{s,l}} \right), \label{F_bs_bar_theta_theta_2}
	\end{align}}where $\mathbf{C}_{\theta_{l},l} = {\dot{{\mathbf{a}}}_{\theta_{l}} {\mathbf{a}} _l^T + {{ {\mathbf{a}} }_l}{\dot{{\mathbf{a}}}_{\theta_{l}}^T}}$. As such, we obtain ${\mathbf{F}}_{\theta_{l},\theta_l}$.
	Similarly, the closed-form expressions of ${{F}}_{\phi_{l},\phi_l}$, ${{F}}_{\theta_{l},\phi_l}$, ${{F}}_{\theta_{l},\bm{\beta}_{l}}$, ${{F}}_{\phi_{l},\bm{\beta}_{l}}$, and ${\mathbf{F}}_{{\bm{\beta}_{l}},\bm{\beta}_{l}}$ are obtained in \eqref{F_theta_beta} -\eqref{F_beta_beta},
where $\mathbf{C}_{\phi_{l},l} = \mathbf{G}_{l}^{T}\mathbf{\Psi}_{l} \left({\dot{{\mathbf{a}}}_{\phi_{l}} {\mathbf{a}} _l^T + {{ {\mathbf{a}} }_l}{\dot{{\mathbf{a}}}_{\phi_{l}}^T}}\right) {{\mathbf{\Psi }}_l}{{\mathbf{G}}_l}$ and $\mathbf{H}_{l} =\mathbf{G}_{l}^{T}\mathbf{\Psi}_{l} {{\mathbf{a}}}_{l} {\mathbf{a}} _l^T  {{\mathbf{\Psi }}_l}{{\mathbf{G}}_l}$. As a result, Proposition \ref{prop_FIM_bs} is proved.

	\section{Proof of Lemma \ref{FIM_appr_lem}}\label{FIM_appr_proof}
	Based on the partial derivative $\dot {\mathbf{a}}_{{\theta _l}} = \bm{Z}_{\theta_l} {{\mathbf{a}}_l}$ in \eqref{De_a_theta}  and $\dot {\mathbf{a}}_{{\phi _l}} = \bm{Z}_{\phi_l} {{\mathbf{a}}_l}$ in \eqref{De_a_phi}, respectively, we have
	\begin{align}
		\mathbf{C}_{\theta_{l},l} & = \mathbf{G}_{l}^{T} \mathbf{A}_{l}\left(\bm{Z}_{\theta_l} \bm{\psi}_{l}\bm{\psi}_{l}^{T}+\bm{\psi}_{l}\bm{\psi}_{l}^{T} \bm{Z}_{\theta_l} \right) \mathbf{A}_{l}{{\mathbf{G}}_l}, \\
		\mathbf{C}_{\phi_{l},l}   & = \mathbf{G}_{l}^{T} \mathbf{A}_{l}\left(\bm{Z}_{\phi_l} \bm{\psi}_{l}\bm{\psi}_{l}^{T}+\bm{\psi}_{l}\bm{\psi}_{l}^{T} \bm{Z}_{\phi_l} \right) \mathbf{A}_{l}{{\mathbf{G}}_l},     \\
		\mathbf{H}_{l}            & = \mathbf{G}_{l}^{T}\mathbf{A}_{l}\bm{\psi}_{l}{{\bm{\psi }}_l^{T}}\mathbf{A}_{l}{{\mathbf{G}}_l},
	\end{align}
	where $\mathbf{A}_{l} = \mathrm{diag}(\mathbf{a}_{l})$. By defining $\mathbf{\Theta}_{l} = \bm{\psi}_{l}\bm{\psi}_{l}^{H}$, we transform the trace term in \eqref{F_theta_theta} as
		{\small \begin{align}
				 & Q_{\theta_{l},\theta_l}(\mathbf{\Theta}_{l}) =\mathrm{tr}\left(  \mathbf{C}_{\theta_{l},l}^H {\mathbf{R}_{{\mathbf{w}}_{l}}^{ - 1}} \mathbf{C}_{\theta_{l},l} {\mathbf{R}_{s,l}}\right)\nonumber                                                                                                                                                                                 \\
				 & ={\mathrm{tr}}\left(\! \mathbf{G}_{l}^{H} \mathbf{A}_{l}^{H} \left(\bm{\psi}_{l}^{*} \bm{\psi}_{l}^{H} \bm{Z}_{\theta_l}^{H} + \bm{Z}_{\theta_l}^{H} \bm{\psi}_{l}^{*}\bm{\psi}_{l}^{H}  \right) \mathbf{A}_{l}^{H}{{\mathbf{G}}_l}^{*} {\mathbf{R}_{{\mathbf{w}}_{l}}^{ - 1}} \mathbf{G}_{l}^{T} \mathbf{A}_{l}\times\right.\nonumber                                             \\
				 & \left.\left(\!\bm{Z}_{\theta_l} \bm{\psi}_{l}\bm{\psi}_{l}^{T}+\bm{\psi}_{l}\bm{\psi}_{l}^{T} \bm{Z}_{\theta_l} \right) \mathbf{A}_{l}{{\mathbf{G}}_l} {\mathbf{R}_{s,l}} \right) \nonumber                                                                                                                                                                                        \\
				 & ={\mathrm{tr}}\!\left(\! \mathbf{R}_{1} \mathbf{\Theta}_{l}^{T}\right)\! {\mathrm{tr}}\!\left(\!  \bm{Z}_{\theta_l}^{H}\mathbf{R}_{2}\bm{Z}_{\theta_l} \mathbf{\Theta}_{l} \right)\!+\! {\mathrm{tr}}\!\left(\! \bm{Z}_{\theta_l} \mathbf{R}_{1} \mathbf{\Theta}_{l}^{T}\right){\mathrm{tr}}\! \left(\!\bm{Z}_{\theta_l}^{H}\mathbf{R}_{2}\mathbf{\Theta}_{l}\right)\nonumber                          \\
				 & + {\mathrm{tr}}\!\left(\!\mathbf{R}_{1} \bm{Z}_{\theta_l}^{H}\mathbf{\Theta}_{l}^{T}\right)\! {\mathrm{tr}}\!\left(\!  \mathbf{R}_{2}\bm{Z}_{\theta_l}\mathbf{\Theta}_{l} \right)  \!+\! {\mathrm{tr}}\!\left(\! \bm{Z}_{\theta_l} \mathbf{R}_{1} \bm{Z}_{\theta_l}^{H} \mathbf{\Theta}_{l}^{T} \right)\!{\mathrm{tr}}\! \left( \! \mathbf{R}_{2} \mathbf{\Theta}_{l}\right),\label{Q_theta_theta}
			\end{align}}where $\mathbf{R}_{1} = \mathbf{A}_{l}{{\mathbf{G}}_l} {\mathbf{R}_{s,l}} \mathbf{G}_{l}^{H} \mathbf{A}_{l}^{H}$ and $\mathbf{R}_{2} = \mathbf{A}_{l}^{H}{{\mathbf{G}}_l}^{*} {\mathbf{R}_{{\mathbf{w}}_{l}}^{ - 1}} \mathbf{G}_{l}^{T} \mathbf{A}_{l}$. Similar to \eqref{Q_theta_theta}, the trace terms in \eqref{F_theta_theta}-\eqref{F_theta_beta} are transformed into
			{\small
			\begin{align}
				 & Q_{\phi_{l},\phi_l}(\mathbf{\Theta}_{l}) = \mathrm{tr}\left(  \mathbf{C}_{\phi_{l},l}^H {\mathbf{R}_{{\mathbf{w}}_{l}}^{ - 1}} \mathbf{C}_{\phi_{l},l} {\mathbf{R}_{s,l}}\right) = \nonumber\\
				 &{\mathrm{tr}}\left(\! \mathbf{R}_{1} \mathbf{\Theta}_{l}^{T}\right)\! {\mathrm{tr}}\left(\! \bm{Z}_{\phi_l}^{H}\mathbf{R}_{2}\bm{Z}_{\phi_l} \mathbf{\Theta}_{l} \right) \!+\! {\mathrm{tr}}\left(\! \bm{Z}_{\phi_l} \mathbf{R}_{1} \mathbf{\Theta}_{l}^{T}\right){\mathrm{tr}} \left(\!\bm{Z}_{\phi_l}^{H}\mathbf{R}_{2}\mathbf{\Theta}_{l}\right) \nonumber \\
				 & \!+\! {\mathrm{tr}}\!\left(\!\mathbf{R}_{1} \bm{Z}_{\phi_l}^{H}\mathbf{\Theta}_{l}^{T}\right)\! {\mathrm{tr}}\left(  \mathbf{R}_{2}\bm{Z}_{\phi_l}\mathbf{\Theta}_{l} \right) \!+\! {\mathrm{tr}}\!\left(\! \bm{Z}_{\phi_l} \mathbf{R}_{1} \bm{Z}_{\phi_l}^{H} \mathbf{\Theta}_{l}^{T} \right)\!{\mathrm{tr}} \left(  \mathbf{R}_{2} \mathbf{\Theta}_{l}\right),\label{Q_phi_phi} \\
				 & Q_{\theta_{l},\phi_l}(\mathbf{\Theta}_{l}) = \mathrm{tr}\left(  \mathbf{C}_{\theta_{l},l}^H {\mathbf{R}_{{\mathbf{w}}_{l}}^{ - 1}} \mathbf{C}_{\phi_{l},l} {\mathbf{R}_{s,l}}\right) = \nonumber\\
				 &{\mathrm{tr}}\!\left(\! \mathbf{R}_{1} \mathbf{\Theta}_{l}^{T}\right) {\mathrm{tr}}\!\left(\!  \bm{Z}_{\theta_l}^{H}\mathbf{R}_{2}\bm{Z}_{\phi_l} \mathbf{\Theta}_{l} \right) \!+\! {\mathrm{tr}}\!\left(\! \bm{Z}_{\phi_l} \mathbf{R}_{1} \mathbf{\Theta}_{l}^{T}\right){\mathrm{tr}}\! \left(\! \bm{Z}_{\theta_l}^{H}\mathbf{R}_{2}\mathbf{\Theta}_{l}\right)\nonumber \\
				 & \!+\! {\mathrm{tr}}\!\left(\!\mathbf{R}_{1} \bm{Z}_{\theta_l}^{H}\mathbf{\Theta}_{l}^{T}\right)\! {\mathrm{tr}}\left(\!  \mathbf{R}_{2}\bm{Z}_{\phi_l}\mathbf{\Theta}_{l} \right) \!+\! {\mathrm{tr}}\left(\! \bm{Z}_{\phi_l} \mathbf{R}_{1} \bm{Z}_{\theta_l}^{H} \mathbf{\Theta}_{l}^{T} \right){\mathrm{tr}}\! \left(\! \mathbf{R}_{2} \mathbf{\Theta}_{l}\right),\label{Q_theta_phi} \\
				 & Q_{\theta_{l},\bm{\beta}_{l}}(\mathbf{\Theta}_{l})= \mathrm{tr}\left( \mathbf{C}_{\theta_{l},l}^H  {\mathbf{R}_{{\mathbf{w}}_{l}}^{ - 1}}\mathbf{H}_{l} {\mathbf{R}_{s,l}} \right) = \nonumber\\
				 &{\mathrm{tr}}\left(\mathbf{R}_{1} \mathbf{\Theta}_{l}^{T}\right) {\mathrm{tr}}\left(\bm{Z}_{\theta_l}^{H} \mathbf{R}_{2}\mathbf{\Theta}_{l} \right)  + {\mathrm{tr}}\left(\mathbf{R}_{1} \bm{Z}_{\theta_l}^{H} \mathbf{\Theta}_{l}^{T}\right) {\mathrm{tr}}\left( \mathbf{R}_{2}\mathbf{\Theta}_{l} \right), \label{Q_theta_beta}  \\
				 & Q_{\phi_{l},\bm{\beta}_{l}}(\mathbf{\Theta}_{l})  = \mathrm{tr}\left(\mathbf{C}_{\phi_{l},l}^H   {\mathbf{R}_{{\mathbf{w}}_{l}}^{ - 1}} \mathbf{H}_{l} {\mathbf{R}_{s,l}} \right) = \nonumber\\
				 &{\mathrm{tr}}\left(\mathbf{R}_{1} \mathbf{\Theta}_{l}^{T}\right) {\mathrm{tr}}\left(\bm{Z}_{\phi_l}^{H} \mathbf{R}_{2}\mathbf{\Theta}_{l} \right) + {\mathrm{tr}}\left(\mathbf{R}_{1} \bm{Z}_{\phi_l}^{H} \mathbf{\Theta}_{l}^{T}\right) {\mathrm{tr}}\left( \mathbf{R}_{2}\mathbf{\Theta}_{l} \right), \label{F_phi_beta_trans} \\
				 & Q_{{\bm{\beta}_{l}},\bm{\beta}_{l}}(\mathbf{\Theta}_{l}) \!=\! \mathrm{tr}\!\left(\!  \mathbf{H}_{l}^H   {\mathbf{R}_{{\mathbf{w}}_{l}}^{ - 1}} \mathbf{H}_{l} {\mathbf{R}_{s,l}} \right) \!=\! {\mathrm{tr}}\!\left(\mathbf{R}_{1} \mathbf{\Theta}_{l}^{T}\right)\! {\mathrm{tr}}\!\left(\!\mathbf{R}_{2}\mathbf{\Theta}_{l} \right),\label{Q_beta_beta}
			\end{align}}respectively. Then, we assume ${\mathbf{R}_{{\mathbf{w}}_{l}}^{ - 1}}$ is constant and derive the derivatives of \eqref{Q_theta_theta}-\eqref{Q_beta_beta} with respect to $\mathbf{\Theta}_{l}$ as
	{\small \begin{align}
		 & \nabla_{\theta_{l},\theta_l}(\mathbf{\Theta}_{l})  = \frac{\partial}{\partial \mathbf{\Theta}_{l}}\mathrm{tr}\left(  \mathbf{C}_{\theta_{l},l}^H {\mathbf{R}_{{\mathbf{w}}_{l}}^{ - 1}} \mathbf{C}_{\theta_{l},l} {\mathbf{R}_{s,l}}\right)\nonumber                \\
		 & ={\mathrm{tr}}\left(  \bm{Z}_{\theta_l}^{H}\mathbf{R}_{2}\bm{Z}_{\theta_l} \mathbf{\Theta}_{l} \right)\mathbf{R}_{1} \!+\! {\mathrm{tr}}\left( \mathbf{R}_{1} \mathbf{\Theta}_{l}^{T}\right) \bm{Z}_{\theta_l}^{T}\mathbf{R}_{2}^{T}\bm{Z}_{\theta_l}^{*} \nonumber \\
		 & + {\mathrm{tr}} \left(\bm{Z}_{\theta_l}^{H}\mathbf{R}_{2}\mathbf{\Theta}_{l}\right)\bm{Z}_{\theta_l} \mathbf{R}_{1} \! + \! {\mathrm{tr}}\left( \bm{Z}_{\theta_l} \mathbf{R}_{1} \mathbf{\Theta}_{l}^{T}\right)\mathbf{R}_{2}^{T} \bm{Z}_{\theta_l}^{*} \nonumber   \\
		 & + {\mathrm{tr}}\left(  \mathbf{R}_{2}\bm{Z}_{\theta_l}\mathbf{\Theta}_{l} \right)\mathbf{R}_{1} \bm{Z}_{\theta_l}^{H} \!+\! {\mathrm{tr}}\left(\mathbf{R}_{1} \bm{Z}_{\theta_l}^{H}\mathbf{\Theta}_{l}^{T}\right)\bm{Z}_{\theta_l}^{T} \mathbf{R}_{2}^{T} \nonumber \\
		 & + {\mathrm{tr}} \left(  \mathbf{R}_{2} \mathbf{\Theta}_{l}\right)\bm{Z}_{\theta_l} \mathbf{R}_{1} \bm{Z}_{\theta_l}^{H}  \!+\! {\mathrm{tr}}\left( \bm{Z}_{\theta_l} \mathbf{R}_{1} \bm{Z}_{\theta_l}^{H} \mathbf{\Theta}_{l}^{T} \right)\mathbf{R}_{2}^{T}.\\
		 & \nabla_{\phi_{l},\phi_l}(\mathbf{\Theta}_{l}) = \frac{\partial}{\partial \mathbf{\Theta}_{l}} \mathrm{tr}\left(  \mathbf{C}_{\phi_{l},l}^H {\mathbf{R}_{{\mathbf{w}}_{l}}^{ - 1}} \mathbf{C}_{\phi_{l},l} {\mathbf{R}_{s,l}}\right) \nonumber               \\
		 & ={\mathrm{tr}}\left(  \bm{Z}_{\phi_l}^{H}\mathbf{R}_{2}\bm{Z}_{\phi_l} \mathbf{\Theta}_{l} \right)\mathbf{R}_{1}+ {\mathrm{tr}}\left( \mathbf{R}_{1} \mathbf{\Theta}_{l}^{T}\right) \bm{Z}_{\phi_l}^{T}\mathbf{R}_{2}^{T}\bm{Z}_{\phi_l}^{*} \nonumber      \\
		 & + {\mathrm{tr}} \left(\bm{Z}_{\phi_l}^{H}\mathbf{R}_{2}\mathbf{\Theta}_{l}\right)\bm{Z}_{\phi_l} \mathbf{R}_{1}  + {\mathrm{tr}}\left( \bm{Z}_{\phi_l} \mathbf{R}_{1} \mathbf{\Theta}_{l}^{T}\right)\mathbf{R}_{2}^{T} \bm{Z}_{\phi_l}^{*} \nonumber        \\
		 & + {\mathrm{tr}}\left(  \mathbf{R}_{2}\bm{Z}_{\phi_l}\mathbf{\Theta}_{l} \right)\mathbf{R}_{1} \bm{Z}_{\phi_l}^{H} + {\mathrm{tr}}\left(\mathbf{R}_{1} \bm{Z}_{\phi_l}^{H}\mathbf{\Theta}_{l}^{T}\right)\bm{Z}_{\phi_l}^{T} \mathbf{R}_{2}^{T} \nonumber     \\
		 & + {\mathrm{tr}} \left(  \mathbf{R}_{2} \mathbf{\Theta}_{l}\right)\bm{Z}_{\phi_l} \mathbf{R}_{1} \bm{Z}_{\phi_l}^{H}  + {\mathrm{tr}}\left( \bm{Z}_{\phi_l} \mathbf{R}_{1} \bm{Z}_{\phi_l}^{H} \mathbf{\Theta}_{l}^{T} \right)\mathbf{R}_{2}^{T},            \\
		 & \nabla_{\theta_{l},\phi_l}(\mathbf{\Theta}_{l}) = \frac{\partial}{\partial \mathbf{\Theta}_{l}} \mathrm{tr}\left(  \mathbf{C}_{\theta_{l},l}^H {\mathbf{R}_{{\mathbf{w}}_{l}}^{ - 1}} \mathbf{C}_{\phi_{l},l} {\mathbf{R}_{s,l}}\right) \nonumber           \\
		 & ={\mathrm{tr}}\left(  \bm{Z}_{\theta_l}^{H}\mathbf{R}_{2}\bm{Z}_{\phi_l} \mathbf{\Theta}_{l} \right)\mathbf{R}_{1} + {\mathrm{tr}}\left( \mathbf{R}_{1} \mathbf{\Theta}_{l}^{T}\right) \bm{Z}_{\phi_l}^{T}\mathbf{R}_{2}^{T}\bm{Z}_{\theta_l}^{*} \nonumber \\
		 & + {\mathrm{tr}} \left(\bm{Z}_{\theta_l}^{H}\mathbf{R}_{2}\mathbf{\Theta}_{l}\right)\bm{Z}_{\phi_l} \mathbf{R}_{1}  + {\mathrm{tr}}\left( \bm{Z}_{\phi_l} \mathbf{R}_{1} \mathbf{\Theta}_{l}^{T}\right)\mathbf{R}_{2}^{T} \bm{Z}_{\theta_l}^{*} \nonumber    \\
		 & + {\mathrm{tr}}\left(  \mathbf{R}_{2}\bm{Z}_{\phi_l}\mathbf{\Theta}_{l} \right)\mathbf{R}_{1} \bm{Z}_{\theta_l}^{H} + {\mathrm{tr}}\left(\mathbf{R}_{1} \bm{Z}_{\theta_l}^{H}\mathbf{\Theta}_{l}^{T}\right)\bm{Z}_{\phi_l}^{T} \mathbf{R}_{2}^{T} \nonumber \\
		 & + {\mathrm{tr}} \left(  \mathbf{R}_{2} \mathbf{\Theta}_{l}\right)\bm{Z}_{\phi_l} \mathbf{R}_{1} \bm{Z}_{\theta_l}^{H}  + {\mathrm{tr}}\left( \bm{Z}_{\phi_l} \mathbf{R}_{1} \bm{Z}_{\theta_l}^{H} \mathbf{\Theta}_{l}^{T} \right)\mathbf{R}_{2}^{T},        \\
		 & \nabla_{\theta_{l},\bm{\beta}_{l}}(\mathbf{\Theta}_{l})  = \frac{\partial}{\partial \mathbf{\Theta}_{l}} \mathrm{tr}\left( \mathbf{C}_{\theta_{l},l}^H  {\mathbf{R}_{{\mathbf{w}}_{l}}^{ - 1}}\mathbf{H}_{l} {\mathbf{R}_{s,l}} \right) \nonumber           \\
		 & = {\mathrm{tr}}\left(\bm{Z}_{\theta_l}^{H} \mathbf{R}_{2}\mathbf{\Theta}_{l} \right)\mathbf{R}_{1}  + {\mathrm{tr}}\left(\mathbf{R}_{1} \mathbf{\Theta}_{l}^{T}\right) \mathbf{R}_{2}^{T} \bm{Z}_{\theta_l}^{*}\nonumber                                    \\
		 & + {\mathrm{tr}}\left( \mathbf{R}_{2}\mathbf{\Theta}_{l} \right)\mathbf{R}_{1} \bm{Z}_{\theta_l}^{H}  + {\mathrm{tr}}\left(\mathbf{R}_{1} \bm{Z}_{\theta_l}^{H} \mathbf{\Theta}_{l}^{T}\right) \mathbf{R}_{2}^{T},                                           \\
		 & \nabla_{\phi_{l},\bm{\beta}_{l}}(\mathbf{\Theta}_{l}) = \frac{\partial}{\partial \mathbf{\Theta}_{l}} \mathrm{tr}\left( \mathbf{C}_{\phi_{l},l}^H  {\mathbf{R}_{{\mathbf{w}}_{l}}^{ - 1}}\mathbf{H}_{l} {\mathbf{R}_{s,l}} \right) \nonumber                \\
		 & = {\mathrm{tr}}\left(\bm{Z}_{\phi_l}^{H} \mathbf{R}_{2}\mathbf{\Theta}_{l} \right)\mathbf{R}_{1}  + {\mathrm{tr}}\left(\mathbf{R}_{1} \mathbf{\Theta}_{l}^{T}\right) \mathbf{R}_{2}^{T} \bm{Z}_{\phi_l}^{*}\nonumber                                        \\
		 & + {\mathrm{tr}}\left( \mathbf{R}_{2}\mathbf{\Theta}_{l} \right)\mathbf{R}_{1} \bm{Z}_{\phi_l}^{H}  + {\mathrm{tr}}\left(\mathbf{R}_{1} \bm{Z}_{\phi_l}^{H} \mathbf{\Theta}_{l}^{T}\right) \mathbf{R}_{2}^{T},                                               \\
		 & \nabla_{{\bm{\beta}_{l}},\bm{\beta}_{l}}(\mathbf{\Theta}_{l}) = \frac{\partial}{\partial \mathbf{\Theta}_{l}} \mathrm{tr}\left(  \mathbf{H}_{l}^H   {\mathbf{R}_{{\mathbf{w}}_{l}}^{ - 1}} \mathbf{H}_{l} {\mathbf{R}_{s,l}} \right)\nonumber               \\
		 & = {\mathrm{tr}}\left(\mathbf{R}_{2}\mathbf{\Theta}_{l} \right)\mathbf{R}_{1} + {\mathrm{tr}}\left(\mathbf{R}_{1} \mathbf{\Theta}_{l}^{T}\right) \mathbf{R}_{2}^{T}.
	\end{align}}As a result, Lemma \ref{FIM_appr_lem} is proved.

	\section{Proof of Proposition \ref{prop_FIM2}}\label{prop_FIM2_proof}
	The derivative of $\bar{\mathbf{a}}_{l}$ with respect to $\theta_{l}$ is given by
	\begin{align}
		 & \frac{{\partial {\bar{\mathbf{a}}_l}}}{{\partial {\theta _l}}} = \frac{{\partial {{\bar{\mathbf{a}}}_v}({\theta _l}) \otimes {{\bar{\mathbf{a}}}_h}({\theta _l},{\phi _l})}}{{\partial {\theta _l}}}\nonumber                                                                     \\
		 & = \frac{{\partial {{\bar{\mathbf{a}}}_v}({\theta _l})}}{{\partial {\theta _l}}} \otimes {\bar{\mathbf{a}}_h}({\theta _l},{\phi _l}) + {\bar{\mathbf{a}}_v}({\theta _l}) \otimes \frac{{\partial {{\bar{\mathbf{a}}}_h}({\theta _l},{\phi _l})}}{{ \partial {\theta _l}}}\nonumber \\
		 & =  - j\frac{{2\pi {{\bar d}_v}}}{\lambda }\sin ({\theta _l})\left( {{\mathbf{d}_{{{\bar N}_v}}} \otimes {\mathbf{1}_{{{\bar N}_h}}}} \right) \circ {\bar{\mathbf{a}}_l} \nonumber                                                                                                 \\
		 & + j\frac{{2\pi {{\bar d}_h}}}{\lambda }\cos ({\theta _l})\cos ({\phi _l})\left( {{\mathbf{1}_{{{\bar N}_v}}}\! \otimes {\mathbf{d}_{{{\bar N}_h}}}} \right) \circ  {\bar{\mathbf{a}}_l} \nonumber                                                                                 \\
		 & = \bar{\bm{\zeta}}_{\theta_l} \circ {\bar{\mathbf{a}}_l} = \bar{\bm{Z}}_{\theta_l} {\bar{\mathbf{a}}_l},\label{De_bar_a_theta}
	\end{align}
	where $ \bar{\bm{Z}}_{\theta_l} = \mathrm{diag}(\bar{\bm{\zeta}}_{\theta_l})$ with $\bar{\bm{\zeta}}_{\theta_l}\! =\! j\frac{{2\pi {{\bar d}_h}}}{\lambda }\cos ({\theta _l})\cos ({\phi _l})\left( {{\mathbf{1}_{{{\bar N}_v}}} \otimes {\mathbf{d}_{{{\bar N}_h}}}} \right) \!-\! j\frac{{2\pi {{\bar d}_v}}}{\lambda }\sin ({\theta _l})\left( {{\mathbf{d}_{{{\bar N}_v}}} \otimes {\mathbf{1}_{{{\bar N}_h}}}} \right)$. Similarly, the derivative of $\bar{\mathbf{a}}_{l}$ with respect to $\phi_{l}$ is
	\begin{align}
		 & \frac{{\partial {\bar{\mathbf{a}}_l}}}{{\partial {\phi_l}}} = \frac{{\partial {{\bar{\mathbf{a}}}_v}({\theta _l}) \otimes {{\bar{\mathbf{a}}}_h}({\theta _l},{\phi _l})}}{{\partial {\phi _l}}}\nonumber                                                                      \\
		 & = \frac{{\partial {{\bar{\mathbf{a}}}_v}({\theta _l})}}{{\partial {\phi _l}}} \otimes {\bar{\mathbf{a}}_h}({\theta _l},{\phi _l}) + {\bar{\mathbf{a}}_v}({\theta _l}) \otimes \frac{{\partial {{\bar{\mathbf{a}}}_h}({\theta _l},{\phi _l})}}{{ \partial {\phi _l}}}\nonumber \\
		 & = \bar{\bm{\zeta}}_{\phi_l} \circ {\bar{\mathbf{a}}_l} = \bar{\bm{Z}}_{\phi_l} {\bar{\mathbf{a}}_l},\label{De_bar_a_phi}
	\end{align}
	where $\bar{\bm{Z}}_{\phi_l} = \mathrm{diag}(\bar{\bm{\zeta}}_{\phi_l})$ with $\bar{\bm{\zeta}}_{\phi_l} =- j\frac{{2\pi {\bar{ d}_h}}}{\lambda }\sin ({\theta _l})\sin ({\phi _l})\left( {{\mathbf{1}_{{{ N}_v}}} \otimes {\mathbf{d}_{{\bar{ N}_h}}}} \right)$.
	%

	First, we derive $\bar{{F}}_{\theta_{l},\theta_l}$, $\bar{{F}}_{\phi_{l},\phi_l}$, and $\bar{\mathbf{F}}_{{\bm{\beta}_{l}},\bm{\beta}_{l}}$.  $\bar{\mathbf{F}}_{\theta_{l},\theta_l}$ is given by
	{\small \begin{align}
		\!\!\!\bar{\mathbf{F}}_{\theta_{l},\theta_l} \!=\! \mathrm{tr}\left(\mathbf{R}_{\bar{\mathbf{y}}_{l}}^{-1}\frac{\partial \mathbf{R}_{\bar{\mathbf{y}}_{l}}}{\partial \theta_{l}}\mathbf{R}_{\bar{\mathbf{y}}_{l}}^{-1}\frac{\partial \mathbf{R}_{\bar{\mathbf{y}}_{l}}}{\partial \theta_{l}}\right) \!\!+ \! 2\Re\left(\frac{\partial \bar{\bm{\eta}}_{l}^{H}}{\partial \theta_{l} } \mathbf{R}_{\bar{\mathbf{y}}_{l}}^{-1} \frac{\partial \bar{\bm{\eta}}_{l}}{\partial \theta_{l} }\right). \label{FIM_theta_theta}
	\end{align}}Note that $\mathbf{R}_{\bar{\mathbf{w}}_{l}}$ is rewritten as $\mathbf{R}_{\bar{\mathbf{w}}_{l}} = \sigma_{\text{r}}^{2}\bar{\mathbf{E}}_{l}\mathbf{\Psi}_{l}\mathbf{\Psi}_{l}^{H}\bar{\mathbf{E}}_{l}^{H} + \sigma_{\text{s}}^{2}\mathbf{I}_{\bar{N}} = \sigma_{\text{r}}^{2}|\beta_{l}|^{2}\mathrm{tr}\left(\mathbf{\Theta}_{l}\right)\bar{\mathbf{a}}_{l}\bar{\mathbf{a}}_{l}^{H} + \sigma_{\text{s}}^{2}\mathbf{I}_{\bar{N}}$. Then, by substituting \eqref{R_y_bar} into \eqref{FIM_theta_theta}, we have
	{\small \begin{align}
		 & {\mathrm{tr}}\left( {{\mathbf{R}}_{{{\bar {\mathbf{y}} }_l}}^{ - 1}\frac{{\partial {{\mathbf{R}}_{{{\bar {\mathbf{y}} }_l}}}}}{{\partial {\theta _l}}}{\mathbf{R}}_{{{\bar {\mathbf{y}} }_l}}^{ - 1}\frac{{\partial {{\mathbf{R}}_{{{\bar {\mathbf{y}} }_l}}}}}{{\partial {\theta _l}}}} \right) = \nonumber                                                                                                                                        \\
		 &  \sigma _r^4{\frac{T_{c}}{L}}\left(\mathrm{tr}\left(\mathbf{\Theta}_{l}\right)\right)^{2}{\mathrm{tr}}\left( {\mathbf{R}_{\bar{\mathbf{w}}_{l}}^{ - 1}}\frac{{\partial |\beta_{l}|^{2}\bar{\mathbf{a}}_{l}\bar{\mathbf{a}}_{l}^{H} }}{{\partial {\theta _l}}}{\mathbf{R}_{\bar{\mathbf{w}}_{l}}^{ - 1}}\frac{{\partial |\beta_{l}|^{2}\bar{\mathbf{a}}_{l}\bar{\mathbf{a}}_{l}^{H}}}{{\partial {\theta _l}}} \right).\label{F_bar_theta_theta_1}
	\end{align}}
	By using \eqref{De_bar_a_theta}, the derivative term $\frac{{\partial \bar{\mathbf{a}}_{l}\bar{\mathbf{a}}_{l}^{H} }}{{\partial {\theta _l}}}$ is obtained as
	\begin{align}
		 & \frac{{\partial \bar{\mathbf{a}}_{l}\bar{\mathbf{a}}_{l}^{H} }}{{\partial {\theta _l}}} =  {{\dot{\bar{\mathbf{a}}}_{\theta_{l}}}\bar{\mathbf{a}}_l^H + {\bar{\mathbf{a}}_l}\dot{\bar{\mathbf{a}}}_{\theta_{l}}^H} ,\label{de_EE_bar_H}
	\end{align}
	where $\dot{\bar{\mathbf{a}}}_{\theta_{l}} = \frac{{\partial {\bar{\mathbf{a}}_l}}}{{\partial {\theta _l}}} = \bar{\bm{\zeta}}_{\theta_l} \circ {\bar{\mathbf{a}}_l} = \bar{\bm{Z}}_{\theta_l} {\bar{\mathbf{a}}_l}$. Then, by substituting \eqref{de_EE_bar_H} into \eqref{F_bar_theta_theta_1}, we have
	\begin{align}
		 & {\mathrm{tr}}\left( {{\mathbf{R}}_{{{\bar {\mathbf{y}} }_l}}^{ - 1}\frac{{\partial {{\mathbf{R}}_{{{\bar {\mathbf{y}} }_l}}}}}{{\partial {\theta _l}}}{\mathbf{R}}_{{{\bar {\mathbf{y}} }_l}}^{ - 1}\frac{{\partial {{\mathbf{R}}_{{{\bar {\mathbf{y}} }_l}}}}}{{\partial {\theta _l}}}} \right)  = \nonumber                          \\
		 & {\frac{4T_{c}}{L}}\sigma _r^4{\left| {{\beta _l}} \right|^4} \left(\mathrm{tr}\left(\mathbf{\Theta}_{l}\right)\right)^{2} {\mathrm{tr}}\left( {\mathbf{R}_{\bar{\mathbf{w}}_{l}}^{ - 1}}\bar{\mathbf{B}}_{\theta_{l},l} {\mathbf{R}_{\bar{\mathbf{w}}_{l}}^{ - 1}}\bar{\mathbf{B}}_{\theta_{l},l}\right),\label{FIM_theta_theta_first}
	\end{align}
	where $\bar{\mathbf{B}}_{\theta_{l},l} = {{\dot{\bar{\mathbf{a}}}_{\theta_{l}}}\bar{\mathbf{a}}_l^H + {\bar{\mathbf{a}}_l}\dot{\bar{\mathbf{a}}}_{\theta_{l}}^H}$. By substituting \eqref{vec_sig_irs} into \eqref{FIM_theta_theta}, we have
	\begin{align}
		 & \Re\left(\frac{\partial \bar{\bm{\eta}}_{l}^{H}}{\partial \theta_{l} } \mathbf{R}_{\bar{\mathbf{y}}_{l}}^{-1} \frac{\partial \bar{\bm{\eta}}_{l}}{\partial \theta_{l} }\right) \nonumber                                                                                                                                                                                                                              \\
		 & =  \Re \left( {\sum\limits_{t = \frac{(l - 1)T_c}{L} + 1}^{l{\frac{T_c}{L}}} {{{\left( {\frac{{\partial {{\bar {\mathbf{E}} }_l}{{\mathbf{\Psi }}_l}{{\mathbf{G}}_l}{{\mathbf{s}}_l}[t]}}{{\partial {\theta _l}}}} \right)}^H}{\mathbf{R}_{\bar{\mathbf{w}}_{l}}^{ - 1}}\frac{{\partial {{\bar {\mathbf{E}} }_l}{{\mathbf{\Psi }}_l}{{\mathbf{G}}_l}{{\mathbf{s}}_l}[t]}}{{\partial {\theta _l}}}} } \right)\nonumber \\
		 & ={\left| {{\beta _l}} \right|^2}\Re \left( \sum\limits_{t = \frac{(l - 1)T_c}{L} + 1}^{l{\frac{T_c}{L}}}  {\mathbf{s}}_l^H[t]{\mathbf{G}}_l^H{\mathbf{\Psi }}_l^H{{\left( {\dot{\bar{\mathbf{a}}}_{\theta_{l}} {\mathbf{a}} _l^T + {{\bar {\mathbf{a}} }_l}{\dot{{\mathbf{a}}}_{\theta_{l}}^T}} \right)}^H} \times \right.\nonumber                                                                                           \\
		 & \left.{\mathbf{R}_{\bar{\mathbf{w}}_{l}}^{ - 1}}\left( {\dot{\bar{\mathbf{a}}}_{\theta_{l}} {\mathbf{a}} _l^T + {{\bar {\mathbf{a}} }_l}{\dot{{\mathbf{a}}}_{\theta_{l}}^T}} \right){{\mathbf{\Psi }}_l}{{\mathbf{G}}_l}{{\mathbf{s}}_l}[t]  \right)  \nonumber                                                                                                                                                         \\
		 & ={\left| {{\beta _l}} \right|^2}\frac{{{T_c}}}{L}\mathrm{tr}\left(  {\bar{\mathbf{C}}_{\theta_{l},l}^H}{\mathbf{R}_{\bar{\mathbf{w}}_{l}}^{ - 1}}\bar{\mathbf{C}}_{\theta_{l},l} {\mathbf{R}_{s,l}} \right),\label{FIM_theta_theta_second}
	\end{align}
	where $\bar{\mathbf{C}}_{\theta_{l},l} = \left( {\dot{\bar{\mathbf{a}}}_{\theta_{l}} {\mathbf{a}} _l^T + {{\bar {\mathbf{a}} }_l}{\dot{{\mathbf{a}}}_{\theta_{l}}^T}} \right) {{\mathbf{\Psi }}_l}{{\mathbf{G}}_l}$. Thus, by combining \eqref{FIM_theta_theta_first} and \eqref{FIM_theta_theta_second}, we obtain
	\begin{align}
		\bar{{F}}_{\theta_{l},\theta_l} & = {\frac{4T_{c}}{L}}\sigma _r^4{\left| {{\beta _l}} \right|^4} \left(\mathrm{tr}\left(\mathbf{\Theta}_{l}\right)\right)^{2} {\mathrm{tr}}\left( {\mathbf{R}_{\bar{\mathbf{w}}_{l}}^{ - 1}}\bar{\mathbf{B}}_{\theta_{l},l} {\mathbf{R}_{\bar{\mathbf{w}}_{l}}^{ - 1}}\bar{\mathbf{B}}_{\theta_{l},l}\right) \nonumber \\
		                                & + \frac{{{2 T_c}}}{L}{\left| {{\beta _l}} \right|^2}\mathrm{tr}\left(  {\bar{\mathbf{C}}_{\theta_{l},l}^H}{\mathbf{R}_{\bar{\mathbf{w}}_{l}}^{ - 1}}\bar{\mathbf{C}}_{\theta_{l},l} {\mathbf{R}_{s,l}} \right).
	\end{align}
	Similar to the derivation of $\bar{{F}}_{\theta_{l},\theta_l} $, we can obtain $\bar{{F}}_{\theta_{l},\theta_l}$,  $\bar{{F}}_{\phi_{l},\phi_l}$, $\bar{{F}}_{\theta_{l},\phi_{l}}$, $\bar{\mathbf{F}}_{\theta_{l},{\bm{\beta}}_l}$, and $\bar{\mathbf{F}}_{{\bm{\beta}_{l}},\bm{\beta}_{l}}$ in \eqref{F_bar_theta_beta}-\eqref{F_bar_beta_beta},
	where $\bar{\mathbf{B}}_{\phi_{l},l} = {{\dot{\bar{\mathbf{a}}}_{\phi_{l}}}{\mathbf{a}}_l^H + {\bar{\mathbf{a}}_l}\dot{{\mathbf{a}}}_{\phi_{l}}^H}$, $\bar{\mathbf{C}}_{\phi_{l},l} = \left( {\dot{\bar{\mathbf{a}}}_{\phi_{l}} {\mathbf{a}} _l^T + {{\bar {\mathbf{a}} }_l}{\dot{{\mathbf{a}}}_{\phi_{l}}^T}} \right) {{\mathbf{\Psi }}_l}{{\mathbf{G}}_l}$, $\bar{\mathbf{D}}_{l,l} = {\bar{\mathbf{a}}_l}\bar{\mathbf{a}}_l^H$, and $\bar{\mathbf{H}}_{l} = {\bar{\mathbf{a}}}_{l} {\mathbf{a}} _l^T {{\mathbf{\Psi }}_l}{{\mathbf{G}}_l}$ with $\dot{\bar{\mathbf{a}}}_{\phi_{l}} = \frac{{\partial {\bar{\mathbf{a}}_l}}}{{\partial {\phi _l}}} = \bar{\bm{\zeta}}_{\phi_l} \circ {\bar{\mathbf{a}}_l} = \bar{\bm{Z}}_{\phi_l} {\bar{\mathbf{a}}_l}$. As a result, Proposition \ref{prop_FIM2} is proved.
	\section{Proof of Lemma \ref{bar_FIM_appr_lem}}\label{bar_FIM_appr_proof}
	The derivative of $\left(\mathrm{tr}\left(\mathbf{\Theta}_{l}\right)\right)^{2} $ with respect to $\mathbf{\Theta}_{l}$ is derived as
	\begin{align}
		\frac{\partial}{\partial \mathbf{\Theta}_{l} } \left(\mathrm{tr}\left(\mathbf{\Theta}_{l}\right)\right)^{2} = 2 \mathrm{tr}\left(\mathbf{\Theta}_{l}\right) \mathbf{I}_{N}.
	\end{align}
	Then, based on the partial derivative $\dot {\mathbf{a}}_{{\theta _l}} = \bar{\bm{Z}}_{\theta_l} {\bar{\mathbf{a}}_l}$ in \eqref{De_bar_a_theta} and $\dot {\mathbf{a}}_{{\phi _l}} = \bar{\bm{Z}}_{\phi_l} {\bar{\mathbf{a}}_l}$ in \eqref{De_bar_a_phi}, respectively, we have
	\begin{align}
		\bar{\mathbf{C}}_{\theta_{l},l} & = \left(\bar{\bm{Z}}_{\theta_l} {\bar{\mathbf{a}}}_{l} \bm{\psi}_{l}^{T}+ {\bar{\mathbf{a}}}_{l} \bm{\psi}_{l}^{T} \bm{Z}_{\theta_l} \right) \mathbf{A}_{l}{{\mathbf{G}}_l}, \\
		\bar{\mathbf{C}}_{\phi_{l},l}   & = \left(\bar{\bm{Z}}_{\phi_l} {\bar{\mathbf{a}}}_{l} \bm{\psi}_{l}^{T}+ {\bar{\mathbf{a}}}_{l} \bm{\psi}_{l}^{T} \bm{Z}_{\phi_l} \right) \mathbf{A}_{l}{{\mathbf{G}}_l},     \\
		\mathbf{H}_{l}                  & = {\bar{\mathbf{a}}}_{l} {\mathbf{a}} _l^T {{\mathbf{\Psi }}_l}{{\mathbf{G}}_l} = {\bar{\mathbf{a}}}_{l} {{\bm{\psi }}_l^{T}}\mathbf{A}_{l}{{\mathbf{G}}_l}.
	\end{align}
	Based on the above derivations, we can transform the second term of the entries of FIM in \eqref{FIM_irs} into the following.
	{\small
	\begin{align}
		 & \bar{Q}_{\theta_{l},\theta_l}(\mathbf{\Theta}_{l}) = \mathrm{tr}\left(  {\bar{\mathbf{C}}_{\theta_{l},l}^H}{\mathbf{R}_{\bar{\mathbf{w}}_{l}}^{ - 1}} \bar{\mathbf{C}}_{\theta_{l},l} {\mathbf{R}_{s,l}}  \right) \nonumber                                                                                                                    \\
		 & = \mathrm{tr}\left({{\mathbf{G}}_l}^{H} \mathbf{A}_{l}^{H} \left(\bm{\psi}_{l}^{*} {\bar{\mathbf{a}}}_{l}^{H} \bar{\bm{Z}}_{\theta_l}^{H}   + \bm{Z}_{\theta_l}^{H} \bm{\psi}_{l}^{*}  {\bar{\mathbf{a}}}_{l}^{H} \right) {\mathbf{R}_{\bar{\mathbf{w}}_{l}}^{ - 1}} \right. \times \nonumber                                                          \\
		 & \left.\left(\bar{\bm{Z}}_{\theta_l} {\bar{\mathbf{a}}}_{l} \bm{\psi}_{l}^{T}+ {\bar{\mathbf{a}}}_{l} \bm{\psi}_{l}^{T} \bm{Z}_{\theta_l} \right) \mathbf{A}_{l}{{\mathbf{G}}_l} {\mathbf{R}_{s,l}}  \right)\nonumber                                                                                                                           \\
		 & =\mathrm{tr} \left( \bar{\bm{Z}}_{\theta_l}^{H}  {\mathbf{R}_{\bar{\mathbf{w}}_{l}}^{ - 1}} \bar{\bm{Z}}_{\theta_l} {\bar{\mathbf{D}}}_{l,l}\right) \mathrm{tr} \left(\mathbf{R}_{1} \mathbf{\Theta}_{l}^{T}  \right) \nonumber                                                                                                                \\
		 & + \mathrm{tr}\left(\bar{\bm{Z}}_{\theta_l}^{H}   {\mathbf{R}_{\bar{\mathbf{w}}_{l}}^{ - 1}} {\bar{\mathbf{D}}}_{l,l} \right) \mathrm{tr} \left( \bm{Z}_{\theta_l} \mathbf{R}_{1} \mathbf{\Theta}_{l}^{T}   \right)\nonumber                                                                                                                    \\
		 & + \mathrm{tr}\left( {\mathbf{R}_{\bar{\mathbf{w}}_{l}}^{ - 1}} \bar{\bm{Z}}_{\theta_l} {\bar{\mathbf{D}}}_{l,l}\right) \mathrm{tr} \left(   \mathbf{R}_{1} \bm{Z}_{\theta_l}^{H} \mathbf{\Theta}_{l}^{T} \right) \nonumber                                                                                                                     \\
		 & + \mathrm{tr}\left( {\mathbf{R}_{\bar{\mathbf{w}}_{l}}^{ - 1}} {\bar{\mathbf{D}}}_{l,l} \right) \mathrm{tr} \left(  \bm{Z}_{\theta_l} \mathbf{R}_{1} \bm{Z}_{\theta_l}^{H} \mathbf{\Theta}_{l}^{T}   \right). \label{bar_Q_theta_theta}
	\end{align}}Similarly, we have
	{\small
	\begin{align}
		 & \bar{Q}_{\phi_{l},\phi_l}(\mathbf{\Theta}_{l}) = \mathrm{tr}\left(  {\bar{\mathbf{C}}_{\theta_{l},l}^H}{\mathbf{R}_{\bar{\mathbf{w}}_{l}}^{ - 1}} \bar{\mathbf{C}}_{\theta_{l},l} {\mathbf{R}_{s,l}}  \right) \nonumber             \\
		 &= \mathrm{tr} \left( \bar{\bm{Z}}_{\phi_l}^{H}  {\mathbf{R}_{\bar{\mathbf{w}}_{l}}^{ - 1}} \bar{\bm{Z}}_{\phi_l} {\bar{\mathbf{D}}}_{l,l}\right) \mathrm{tr} \left(\mathbf{R}_{1} \mathbf{\Theta}_{l}^{T}  \right) \nonumber         \\
		 & + \mathrm{tr}\left(\bar{\bm{Z}}_{\phi_l}^{H}   {\mathbf{R}_{\bar{\mathbf{w}}_{l}}^{ - 1}} {\bar{\mathbf{D}}}_{l,l} \right) \mathrm{tr} \left( \bm{Z}_{\phi_l} \mathbf{R}_{1} \mathbf{\Theta}_{l}^{T}   \right)\nonumber             \\
		 & + \mathrm{tr}\left( {\mathbf{R}_{\bar{\mathbf{w}}_{l}}^{ - 1}} \bar{\bm{Z}}_{\phi_l} {\bar{\mathbf{D}}}_{l,l}\right) \mathrm{tr} \left(   \mathbf{R}_{1} \bm{Z}_{\phi_l}^{H} \mathbf{\Theta}_{l}^{T} \right) \nonumber              \\
		 & + \mathrm{tr}\left( {\mathbf{R}_{\bar{\mathbf{w}}_{l}}^{ - 1}} {\bar{\mathbf{D}}}_{l,l} \right) \mathrm{tr} \left(  \bm{Z}_{\phi_l} \mathbf{R}_{1} \bm{Z}_{\phi_l}^{H} \mathbf{\Theta}_{l}^{T}   \right),\label{bar_Q_phi_phi}      \\
		 & \bar{Q}_{\theta_{l},\phi_{l}}(\mathbf{\Theta}_{l})  = \mathrm{tr}\left(  {\bar{\mathbf{C}}_{\theta_{l},l}^H}{\mathbf{R}_{\bar{\mathbf{w}}_{l}}^{ - 1}} \bar{\mathbf{C}}_{\phi_{l},l}{\mathbf{R}_{s,l}}  \right) \nonumber           \\
		 & = \mathrm{tr} \left( \bar{\bm{Z}}_{\theta_l}^{H}  {\mathbf{R}_{\bar{\mathbf{w}}_{l}}^{ - 1}} \bar{\bm{Z}}_{\phi_l} {\bar{\mathbf{D}}}_{l,l}\right) \mathrm{tr} \left(\mathbf{R}_{1} \mathbf{\Theta}_{l}^{T}  \right) \nonumber      \\
		 & + \mathrm{tr}\left(\bar{\bm{Z}}_{\theta_l}^{H}   {\mathbf{R}_{\bar{\mathbf{w}}_{l}}^{ - 1}} {\bar{\mathbf{D}}}_{l,l} \right) \mathrm{tr} \left( \bm{Z}_{\phi_l} \mathbf{R}_{1} \mathbf{\Theta}_{l}^{T}   \right)\nonumber           \\
		 & + \mathrm{tr}\left( {\mathbf{R}_{\bar{\mathbf{w}}_{l}}^{ - 1}} \bar{\bm{Z}}_{\phi_l} {\bar{\mathbf{D}}}_{l,l}\right) \mathrm{tr} \left(   \mathbf{R}_{1} \bm{Z}_{\theta_l}^{H} \mathbf{\Theta}_{l}^{T} \right) \nonumber            \\
		 & + \mathrm{tr}\left( {\mathbf{R}_{\bar{\mathbf{w}}_{l}}^{ - 1}} {\bar{\mathbf{D}}}_{l,l} \right) \mathrm{tr} \left(  \bm{Z}_{\phi_l} \mathbf{R}_{1} \bm{Z}_{\theta_l}^{H} \mathbf{\Theta}_{l}^{T}   \right), \label{bar_Q_theta_phi} \\
		 & \bar{Q}_{\theta_{l},\bm{\beta}_{l}}(\mathbf{\Theta}_{l})  = \mathrm{tr}\left({ \bar{\mathbf{C}}_{\theta_{l},l}^H}{\mathbf{R}_{\bar{\mathbf{w}}_{l}}^{ - 1}}\bar{\mathbf{H}}_{l}{\mathbf{R}_{s,l}}  \right) \nonumber                \\
		 & = \mathrm{tr} \left( \bar{\bm{Z}}_{\theta_l}^{H}  {\mathbf{R}_{\bar{\mathbf{w}}_{l}}^{ - 1}} {\bar{\mathbf{D}}}_{l,l}\right) \mathrm{tr} \left(\mathbf{R}_{1} \mathbf{\Theta}_{l}^{T}  \right) \nonumber                            \\
		 & + \mathrm{tr}\left( {\mathbf{R}_{\bar{\mathbf{w}}_{l}}^{ - 1}} {\bar{\mathbf{D}}}_{l,l} \right) \mathrm{tr} \left( \mathbf{R}_{1} \bm{Z}_{\theta_l}^{H} \mathbf{\Theta}_{l}^{T}   \right),\label{bar_Q_theta_beta}                  \\
		 & \bar{Q}_{\phi_{l},\bm{\beta}_{l}}(\mathbf{\Theta}_{l})  = \mathrm{tr}\left({ \bar{\mathbf{C}}_{\phi_{l},l}^H}{\mathbf{R}_{\bar{\mathbf{w}}_{l}}^{ - 1}}\bar{\mathbf{H}}_{l}{\mathbf{R}_{s,l}}  \right) \nonumber                    \\
		 & = \mathrm{tr} \left( \bar{\bm{Z}}_{\phi_l}^{H}  {\mathbf{R}_{\bar{\mathbf{w}}_{l}}^{ - 1}} {\bar{\mathbf{D}}}_{l,l}\right) \mathrm{tr} \left(\mathbf{R}_{1} \mathbf{\Theta}_{l}^{T}  \right) \nonumber                              \\
		 & + \mathrm{tr}\left( {\mathbf{R}_{\bar{\mathbf{w}}_{l}}^{ - 1}} {\bar{\mathbf{D}}}_{l,l} \right) \mathrm{tr} \left( \mathbf{R}_{1} \bm{Z}_{\phi_l}^{H} \mathbf{\Theta}_{l}^{T}   \right),\label{bar_Q_phi_beta}                      \\
		 & \bar{Q}_{{\bm{\beta}_{l}},\bm{\beta}_{l}}(\mathbf{\Theta}_{l})  = \mathrm{tr}\left(  { {\bar{\mathbf{H}}_{l}}^H}{\mathbf{R}_{\bar{\mathbf{w}}_{l}}^{ - 1}}\bar{\mathbf{H}}_{l}{\mathbf{R}_{s,l}}  \right) \nonumber                 \\
		 & = \mathrm{tr}\left( {\mathbf{R}_{\bar{\mathbf{w}}_{l}}^{ - 1}} {\bar{\mathbf{D}}}_{l,l} \right) \mathrm{tr}\left( \mathbf{R}_{1} \mathbf{\Theta}_{l}^{T}  \right). \label{bar_Q_beta_beta}
	\end{align}}
	Then, we assume ${\mathbf{R}_{\bar{\mathbf{w}}_{l}}^{ - 1}}$ is constant. As such, the corresponding derivatives of \eqref{bar_Q_theta_theta}-\eqref{bar_Q_beta_beta} with respect to $\mathbf{\Theta}_{l}$ are given by
	{\small
	\begin{align}
		& \bar{\nabla}_{\theta_{l},\theta_l}(\mathbf{\Theta}_{l})  = \frac{\partial}{\partial \mathbf{\Theta}_{l}}\mathrm{tr}\left(  {\bar{\mathbf{C}}_{\theta_{l},l}^H}{\mathbf{R}_{\bar{\mathbf{w}}_{l}}^{ - 1}} \bar{\mathbf{C}}_{\theta_{l},l} {\mathbf{R}_{s,l}}  \right) \nonumber                                                                                            \\
		 & =\mathrm{tr} \left( \bar{\bm{Z}}_{\theta_l}^{H}  {\mathbf{R}_{\bar{\mathbf{w}}_{l}}^{ - 1}} \bar{\bm{Z}}_{\theta_l} {\bar{\mathbf{D}}}_{l,l}\right) \mathrm{tr}\mathbf{R}_{1}  \!+\! \mathrm{tr}\left(\bar{\bm{Z}}_{\theta_l}^{H}   {\mathbf{R}_{\bar{\mathbf{w}}_{l}}^{ - 1}} {\bar{\mathbf{D}}}_{l,l} \right)\bm{Z}_{\theta_l} \mathbf{R}_{1} \nonumber                     \\
		 & +\! \mathrm{tr}\left( {\mathbf{R}_{\bar{\mathbf{w}}_{l}}^{ - 1}} \bar{\bm{Z}}_{\theta_l} {\bar{\mathbf{D}}}_{l,l}\right) \mathbf{R}_{1} \bm{Z}_{\theta_l}^{H}  \!\!+\! \mathrm{tr}\!\left( {\mathbf{R}_{\bar{\mathbf{w}}_{l}}^{ - 1}} {\bar{\mathbf{D}}}_{l,l} \right) \bm{Z}_{\theta_l} \mathbf{R}_{1} \bm{Z}_{\theta_l}^{H},                                                      \\
		 & \bar{\nabla}_{\phi_{l},\phi_l}(\mathbf{\Theta}_{l})  = \frac{\partial}{\partial \mathbf{\Theta}_{l}}\mathrm{tr}\left(  {\bar{\mathbf{C}}_{\phi_{l},l}^H}{\mathbf{R}_{\bar{\mathbf{w}}_{l}}^{ - 1}} \bar{\mathbf{C}}_{\phi_{l},l} {\mathbf{R}_{s,l}}  \right) \nonumber                                                                                                    \\
		 & =\mathrm{tr} \left( \bar{\bm{Z}}_{\phi_l}^{H}  {\mathbf{R}_{\bar{\mathbf{w}}_{l}}^{ - 1}} \bar{\bm{Z}}_{\phi_l} {\bar{\mathbf{D}}}_{l,l}\right) \mathrm{tr}\mathbf{R}_{1}  \!+\! \mathrm{tr}\left(\bar{\bm{Z}}_{\phi_l}^{H}   {\mathbf{R}_{\bar{\mathbf{w}}_{l}}^{ - 1}} {\bar{\mathbf{D}}}_{l,l} \right)\bm{Z}_{\phi_l} \mathbf{R}_{1} \nonumber                             \\
		 & +\! \mathrm{tr}\!\left( {\mathbf{R}_{\bar{\mathbf{w}}_{l}}^{ - 1}} \bar{\bm{Z}}_{\phi_l} {\bar{\mathbf{D}}}_{l,l}\right) \mathbf{R}_{1} \bm{Z}_{\phi_l}^{H}  \!\!+\! \mathrm{tr}\!\left( {\mathbf{R}_{\bar{\mathbf{w}}_{l}}^{ - 1}} {\bar{\mathbf{D}}}_{l,l} \right) \bm{Z}_{\phi_l} \mathbf{R}_{1} \bm{Z}_{\phi_l}^{H},                                                              \\
		 & \bar{\nabla}_{\theta_{l},\phi_l}(\mathbf{\Theta}_{l}) = \frac{\partial}{\partial \mathbf{\Theta}_{l}} \mathrm{tr}\left(  {\bar{\mathbf{C}}_{\theta_{l},l}^H}{\mathbf{R}_{\bar{\mathbf{w}}_{l}}^{ - 1}} \bar{\mathbf{C}}_{\phi_{l},l}{\mathbf{R}_{s,l}}  \right) \nonumber                                                                                                 \\
		 & = \mathrm{tr} \left( \bar{\bm{Z}}_{\theta_l}^{H}  {\mathbf{R}_{\bar{\mathbf{w}}_{l}}^{ - 1}} \bar{\bm{Z}}_{\phi_l} {\bar{\mathbf{D}}}_{l,l}\right)\mathbf{R}_{1} \!+\! \mathrm{tr}\left(\bar{\bm{Z}}_{\theta_l}^{H}   {\mathbf{R}_{\bar{\mathbf{w}}_{l}}^{ - 1}} {\bar{\mathbf{D}}}_{l,l} \right)\bm{Z}_{\phi_l} \mathbf{R}_{1} \nonumber                                     \\
		 & +\! \mathrm{tr}\!\left( {\mathbf{R}_{\bar{\mathbf{w}}_{l}}^{ - 1}} \bar{\bm{Z}}_{\phi_l} {\bar{\mathbf{D}}}_{l,l}\right) \mathbf{R}_{1} \bm{Z}_{\theta_l}^{H} \!\!+\! \mathrm{tr}\!\left( {\mathbf{R}_{\bar{\mathbf{w}}_{l}}^{ - 1}} {\bar{\mathbf{D}}}_{l,l} \right) \bm{Z}_{\phi_l} \mathbf{R}_{1} \bm{Z}_{\theta_l}^{H},                                                           \\
		 & \bar{\nabla}_{\theta_{l},\bm{\beta}_{l}}(\mathbf{\Theta}_{l})  = \frac{\partial}{\partial \mathbf{\Theta}_{l}} \mathrm{tr}\left({ \bar{\mathbf{C}}_{\theta_{l},l}^H}{\mathbf{R}_{\bar{\mathbf{w}}_{l}}^{ - 1}}\bar{\mathbf{H}}_{l}{\mathbf{R}_{s,l}}  \right) \nonumber                                                                                                   \\
		 & = \mathrm{tr} \left( \bar{\bm{Z}}_{\theta_l}^{H}  {\mathbf{R}_{\bar{\mathbf{w}}_{l}}^{ - 1}} {\bar{\mathbf{D}}}_{l,l}\right) \mathrm{tr}\mathbf{R}_{1}  + \mathrm{tr}\left( {\mathbf{R}_{\bar{\mathbf{w}}_{l}}^{ - 1}} {\bar{\mathbf{D}}}_{l,l} \right) \mathbf{R}_{1} \bm{Z}_{\theta_l}^{H},                                                                             \\
		 & \bar{\nabla}_{\phi_{l},\bm{\beta}_{l}}(\mathbf{\Theta}_{l})  = \frac{\partial}{\partial \mathbf{\Theta}_{l}} \mathrm{tr}\left({ \bar{\mathbf{C}}_{\phi_{l},l}^H}{\mathbf{R}_{\bar{\mathbf{w}}_{l}}^{ - 1}}\bar{\mathbf{H}}_{l}{\mathbf{R}_{s,l}}  \right) \nonumber                                                                                                       \\
		 & = \mathrm{tr} \left( \bar{\bm{Z}}_{\phi_l}^{H}  {\mathbf{R}_{\bar{\mathbf{w}}_{l}}^{ - 1}} {\bar{\mathbf{D}}}_{l,l}\right)\mathbf{R}_{1} + \mathrm{tr}\left( {\mathbf{R}_{\bar{\mathbf{w}}_{l}}^{ - 1}} {\bar{\mathbf{D}}}_{l,l} \right)\mathbf{R}_{1} \bm{Z}_{\phi_l}^{H},                                                                                               \\
		 & \bar{\nabla}_{{\bm{\beta}_{l}},\bm{\beta}_{l}}(\mathbf{\Theta}_{l}) \!=\! \frac{\partial}{\partial \mathbf{\Theta}_{l}} \mathrm{tr}\left(  { {\bar{\mathbf{H}}_{l}}^H}{\mathbf{R}_{\bar{\mathbf{w}}_{l}}^{ - 1}}\bar{\mathbf{H}}_{l}{\mathbf{R}_{s,l}}  \right) \!=\! \mathrm{tr}\left( {\mathbf{R}_{\bar{\mathbf{w}}_{l}}^{ - 1}} {\bar{\mathbf{D}}}_{l,l} \right)\mathbf{R}_{1}.
	\end{align}}As a result, Lemma \ref{bar_FIM_appr_lem} is proved.
\end{appendices}
\bibliographystyle{IEEEtran}
\bibliography{IEEEabrv,myref}

\end{document}